\documentclass[english,12pt]{article}
\usepackage[T1]{fontenc}
\usepackage[utf8]{inputenc}
\usepackage[english]{babel}
\usepackage[width=\textwidth]{caption}
\usepackage{amsmath,amsopn,amssymb,bm, dsfont,mathrsfs, bbold}
\everymath{\displaystyle}
\usepackage{a4wide}
\usepackage{array,float, placeins,flafter, longtable,booktabs,pdflscape}
\usepackage{verbatim}
\usepackage{natbib}
\usepackage{lmodern}
\usepackage{xr}
\usepackage{mathtools}
\usepackage{pdfpages}
\usepackage[letterpaper,top=2cm,bottom=2cm,left=3cm,right=3cm,marginparwidth=1.75cm]{geometry}
\usepackage{graphicx,float}
\usepackage[colorlinks=true, allcolors=blue]{hyperref}
\usepackage{listings}
\usepackage{xcolor}
\usepackage{url}

\makeatletter
\renewcommand\@fnsymbol[1]{\@arabic{#1}} 
\makeatother

\newtheorem{hyp}{Assumption}

\newtheorem{conddes}{Design Restriction}

\newcommand{\R}{\mathbb R}

\newcommand{\ind}[1]{\mathds{1}\left\{#1\right\}}

\newcommand{\eps}{\varepsilon}

\newcommand{\st}[1]{\texttt{#1}}
\newcommand{\pl}{\text{pl}}

\newcommand{\DID}{\text{DID}}

\newcommand{\convd}{\stackrel{d}{\to}}
\newcommand{\hbeta}{\hat{\beta}}
\newcommand{\tbeta}{\tilde{\beta}}

\makeatletter
\renewcommand{\section}{\@startsection{section}{2}{0mm}{-1.5\baselineskip}{1\baselineskip}{\normalfont\large\bfseries}}
\renewcommand{\subsection}{\@startsection{subsection}{2}{0mm}{-1.2\baselineskip}{1\baselineskip}{\normalfont\normalsize\bfseries}}
\renewcommand{\subsubsection}{\@startsection{subsubsection}{3}{0mm}{-0.8\baselineskip}{0.4\baselineskip}{\normalfont\normalsize\itshape}}
\makeatother

\newcommand*{\storecounter}[2]{%
  \edef\@currentlabel{\the\value{#1}}%
  \label{#2}%
}

\date{\today}

\begin{document}

\setcounter{footnote}{0}
\renewcommand{\thefootnote}{\arabic{footnote}}

\title{Using \st{did\_multiplegt\_dyn} to Estimate Event-Study Effects in Complex Designs: Overview, and Four Examples Based on Real Datasets\thanks{Chaisemartin, Ciccia, Knau, Malézieux, and Sow wrote the Stata package. Angotti, Ciccia, Fabre, and Quispe translated it into R. Arboleda and Knau ran the simulations. Li handled the applications. Chaisemartin and D'Haultfoeuille proposed the estimators computed by the commands. Arboleda, Chaisemartin, and Ciccia wrote this paper. We are extremely grateful to Tom Stringham for spotting typos in earlier versions of the code and of this technical documentation. We are also extremely grateful to all users of our commands, whose numerous comments have helped us improve it. This project was funded by the European Union (ERC, REALLYCREDIBLE,GA N°101043899). Views and opinions expressed are those of the authors and do not reflect those of the European Union or the European Research Council Executive Agency. Neither the European Union nor the granting authority can be held responsible for them.}}

\author{Cl\'{e}ment de Chaisemartin\thanks{Sciences Po, clement.dechaisemartin@sciencespo.fr} 
\and 
Diego Ciccia\thanks{Northwestern University, diego.ciccia@kellogg.northwestern.edu} 
\and
Felix Knau\thanks{Ludwig-Maximilians-Universität München, felix.knau@econ.lmu.de} 
\and
M\'{e}litine Mal\'{e}zieux\thanks{Stockholm School of Economics, melitine.malezieux@hhs.se} 
\and
Doulo Sow\thanks{Princeton University, doulosow164@gmail.com}
\and
David Arboleda\thanks{Sciences Po, davidenrique.arboledacarcamo@sciencespo.fr} 
\and
Romain Angotti\thanks{Stanford University, rangotti@stanford.edu} 
\and 
Xavier D'Haultf\oe{}uille\thanks{CREST-ENSAE, xavier.dhaultfoeuille@ensae.fr}
\and
Bingxue Li\thanks{University of Illinois at Urbana Champaign, bingxue5@illinois.edu}
\and
Henri Fabre\thanks{CREST-ENSAE, henri.fabre@ensae.fr}
\and
Anzony Quispe\thanks{Sciences Po, anzony.quisperojas@sciencespo.fr}}

\maketitle
\vspace{-1cm}

\begin{abstract}
The command did\_multiplegt\_dyn can be used to estimate event-study effects in complex designs with a potentially non-binary and/or non-absorbing treatment. This paper starts by providing an overview of the estimators computed by the command. Then, simulations based on three real datasets are used to demonstrate the estimators' properties. Finally, the command is used on four real datasets to estimate event-study effects in complex designs. The first example has a binary treatment that can turn on an off. The second example has a continuous absorbing treatment. The third example has a discrete multivalued treatment that can increase or decrease multiple times over time. The fourth example has two, binary and absorbing treatments, where the second treatment always happens after the first.
\end{abstract}

\newpage

\section{Introduction}

This paper presents the \st{did\_multiplegt\_dyn} Stata and R commands, respectively available from the SSC and CRAN repositories, which compute the event-study estimators proposed by \cite{de2022intertemporal}, hereafter referred to as the paper.  Those estimators can already be computed by the \st{did\_multiplegt} Stata command, when the \st{robust\_dynamic} option is specified. However, \st{did\_multiplegt} relies on the boostrap for variances' estimation, while \st{did\_multiplegt\_dyn} relies on analytic formulas. Thus, \st{did\_multiplegt\_dyn} is much faster than \st{did\_multiplegt}. It is also more closely aligned with the final version of the paper, and it has more options. Therefore, it supersedes \st{did\_multiplegt} for the computation of the estimators proposed by \cite{de2022intertemporal}.

\medskip
The main goal of this companion note is to increase the transparency of the  \st{did\_multiplegt\_dyn} command, by bridging the paper and its ado. For that purpose, it gives explicit formulas for the estimators and the variance estimators produced by the command, with more details than in \cite{de2022intertemporal}. Moreover, for many of the key quantities to compute the estimators and their variances, this companion note gives tags one can use to locate the lines computing those quantities in the .ado and .R files. 

\medskip
This companion note is organized as follows. In Section 1, we review the baseline event-study estimators computed by the command without any option other than \st{effects}. In Section 2, we review other estimators computed by the command when other options are specified. In particular, we outline in details estimators with control variables. Section 3 presents some simulations results. In Appendix A, we explain how the command deals with imbalanced panels. In Appendix B, we review instances where \st{did\_multiplegt} and \st{did\_multiplegt\_dyn} can output different point estimates, and explain the source of those differences.

\section{Baseline estimators: \st{did\_multiplegt\_dyn Y G T D, effects(\#)}}  
\label{sec:baseline}

\subsection{Set-up and identifying assumptions}  
\label{subsec:setup}

\paragraph{Data.} We seek to estimate the effect of a treatment on an outcome. For that purpose, we use a panel of $G$ groups observed at $T$ periods, respectively indexed by $g$ and $t$. For now, we assume that the panel of groups is balanced: each group is observed at every time period.

\paragraph{Weight assigned to cell $(g,t)$.}
Let $N_{g,t}$ denote the weight assigned to cell $(g,t)$ in the estimation. If the data is at the $(g,t)$ level and the $\st{weight}$ option is not specified, $N_{g,t}=1$. If the data is at the $(g,t)$ level and the $\st{weight}$ option is specified, $N_{g,t}$ is equal to the argument inputted to that option. If the data is at a more disaggregated level than the $(g,t)$ level
and the $\st{weight}$ option is not specified, $N_{g,t}$ is the number of observations with non-missing outcome data in cell $(g,t)$. Finally, if the data is at a more disaggregated level than the $(g,t)$ level and the $\st{weight}$ option is specified, $N_{g,t}$ is equal to the summation of the argument inputted to that option across all observations with non-missing outcome data in cell $(g,t)$. \textit{Code tag: see ``Collapse and weight'' section in the code.}

\paragraph{Treatment and outcome definitions.}
If the data is at the $(g,t)$ level, let $D_{g,t}$ and $Y_{g,t}$ denote the treatment and outcome of cell $(g,t)$. If the data is at a more disaggregated level than the $(g,t)$ level, let $D_{g,t}$ and $Y_{g,t}$ denote the average treatment and outcome of all observations with non-missing outcome data in cell $(g,t)$, weighted by the argument of the $\st{weight}$ option if that option is specified. \textit{Code tag: see ``Collapse and weight'' section in the code.}

\paragraph{Potential and observed outcomes.} For all $(d_1,...,d_T)\in \mathcal{D}$, let $Y_{g,t}(d_1,...,d_T)$ denote the potential outcome of group $g$ at $t$ if $(D_{g,1},...,D_{g,T})=(d_1,...,d_T)$, and let $Y_{g,t}=Y_{g,t}(D_{g,1},...,D_{g,T})$ denote the observed outcome of $g$ at $t$. This dynamic potential outcome framework follows \cite{robins1986new}. It explicitly allows groups' outcome at $t$ to depend on their past and future treatments.

\paragraph{Conditioning on the design.} When defining its target parameters, the paper takes the perspective of a social planner, seeking to conduct a cost-benefit analysis comparing groups' actual treatments $\bm{D}\equiv (D_{g,t})_{g\in \{1,...,G\}\times \{1,...,T\}}$  to the counterfactual ``status-quo'' scenario where every group would have kept all the time
the same treatment as in period 1. Then, the analysis is conditional on $\bm{D}$, the study's design. Throughout, we also assume that $N_{g,t}$ is non-stochastic. Then, only groups' potential outcomes are random.

\paragraph{Date of first treatment change.} For all $g$, let $F_g=\min\{t:t\geq 2,D_{g,t}\ne D_{g,t-1}\}$. We adopt the convention that $F_g=T+1$ if $g$'s treatment never changes. \textit{Code tag: see ``Date of first treatment change'' section in the code.} 

\paragraph{Condition that the design has to satisfy for our baseline estimators to be applicable.} 
Our baseline estimators are applicable to any design that satisfies the restriction below.
\begin{conddes}\label{hyp:non_pathological_design}
$\exists (g,g')$ such that: (i) $D_{g,1}=D_{g',1}$, (ii) $F_g\ne F_{g'}$.
\end{conddes}
(i) requires that there exist groups with the same period-one treatment. This is essentially a restriction on the support of the period-one treatment. If groups' period-one treatments are i.i.d. draws from a continuous distribution, $D_{g,1}\ne D_{g',1}$ for all $(g,g')$, so (i) fails. On the other hand, if $D_{g,1}$ can only take a finite number of values $K$, (i) automatically holds as long as $K<G$. When (i) fails, the command can still be used with the $\st{continuous}$ option (see below), but for now we assume that (i) holds. (ii) requires that there is heterogeneity in the date at which groups change treatment for the first time. (ii) for instance fails if groups' treatment is extremely non-persistent, so that $D_{g,1}\ne D_{g,2}$ and $F_g=2$ for all $g$. The command assesses if Design Restriction \ref{hyp:non_pathological_design} is satisfied, and returns an error message if it is not. \textit{Code tag: see ``Error message if Design Restriction 1 is not met'' sections in the code.}

\paragraph{Identifying assumptions}
We maintain throughout the following condition.
\begin{hyp}\label{hyp:no_antic}
	(No Anticipation) $\forall g$, $\forall(d_1,...,d_T)\in \mathcal{D}$, $Y_{g,t}(d_1,...,d_T)=Y_{g,t}(d_1,...,d_t)$.
	\end{hyp}
Assumption \ref{hyp:no_antic} requires that a group's current outcome does not depend on its future treatments, the so-called no-anticipation hypothesis. Let
$\mathcal{D}^{\text{r}}_1=\{d:\exists (g,g')\in\{1,...,G\}^2:D_{g,1}=D_{g',1}=d,F_g\ne F_{g'}\}$
be the set of period-one-treatment values such that two groups with different values of $F_g$ have that period-one treatment. For any $d$ in $\mathcal{D}^{\text{r}}_1$ and any $t$, let $\bm{d}_t$ denote a $1\times t$ vector of $d$s. For any integer $k$, let $\bm{D}_{g,1,k}$ be a $1\times k$ vector whose coordinates are all equal to $D_{g,1}$. $Y_{g,t}(\bm{D}_{g,1,t})$ is $g$'s period-$t$ outcome in a counterfactual where it keeps its period-one treatment till period $t$. We refer to it as its ``status-quo'' potential outcome. In this section, we make the following parallel-trends assumption:
\begin{hyp}\label{hyp:strong_exogeneity}
	(Parallel trends for the status-quo outcome, for groups with the same period-one treatment)
$\forall (g, g’)$, if $D_{g,1} = D_{g’,1} \in \mathcal{D}_1^{\text{r}}$, then $\forall t\geq 2$,
$$E[Y_{g,t}(\bm{D}_{g,1,t}) - Y_{g,t-1}(\bm{D}_{g,1,t-1}) | \bm{D}] = E[Y_{g’,t}(\bm{D}_{g’,1,t}) - Y_{g’,t-1}(\bm{D}_{g’,1,t-1}) | \bm{D}].$$
\end{hyp}
Assumption \ref{hyp:strong_exogeneity} requires that if two groups have the same period-one treatment, then they have the same expected evolution of their status-quo outcome. In other words, if their treatments do not change, all groups with the same period-one treatment have the same expected outcome evolution.

\subsection{Target parameters and estimators}  
\label{subsec:target}

\paragraph{Last period when $g$ still has a control group.} For every $g$, let
$$T_g=\max_{g':D_{g',1}=D_{g,1}}F_{g'}-1$$
denote the last period where there is still a group with the same period-one treatment as $g$ and whose treatment has not changed since the start of the panel. \textit{Code tag: see ``Determining last period where g still has a control group'' section in the code.}

\paragraph{Size of $g$'s control group at $t$.}
For all $(g,t)$, let $N^g_t=\sum_{g':D_{g',1}=D_{g,1},F_{g'} >t}N_{g',t}$ denote the number of observations at period $t$ in groups $g'$ with the same period-one treatment as $g$, and that kept the same treatment from period $1$ to $t$. For every $g$ such that $F_g\leq T_g$, and every $\ell\in \{1,...,T_g-(F_g-1)\}$, we have $N^{g}_{F_g-1+\ell}>0$. \textit{Code tag: see ``number of control groups for g at t'' section in the code.}

\paragraph{AVSQ effect and estimator.}
For any $g$ such that $F_g\leq T_g$, and for any $\ell \in \{1,...,T_g-F_g+1\}$, let
\begin{equation}\label{eq:defAvsSQ}
\delta_{g,\ell}=E\left[Y_{g,F_g-1+\ell}-Y_{g,F_g-1+\ell}(D_{g,1},...,D_{g,1})|\bm{D}\right]
\end{equation}
be the actual-versus-status-quo (AVSQ) effect of $g$ at $F_g-1+\ell$.
Lemma 1 in the paper shows that
\begin{equation}\label{eq:DIDgl}
\DID_{g,\ell} =Y_{g,F_g-1+\ell} - Y_{g,F_g-1} -
\sum_{g':D_{g',1}=D_{g,1},F_{g'} >F_g-1+\ell}\frac{N_{g',F_g-1+\ell}}{N^{g}_{F_g-1+\ell}}(Y_{g',F_g-1+\ell} - Y_{g',F_g-1})
\end{equation}
is unbiased for $\delta_{g,\ell}$.

\paragraph{Defining switchers in and out.}
By default, the command enforces Design Restriction 2 in the paper, meaning that it drops all $(g,t)$ cells such that at $t$, $g$ has experienced a treatment strictly above and strictly below its period-one treatment (\textit{Code tag: see ``Enforcing Design Restriction 2 in the paper'' section in the code}). The paper argues that enforcing this restriction is desirable, as it ensures that the non-normalized event-study effects introduced below can be interpreted as effects of having been exposed to a weakly larger treatment dose for $\ell$ periods. The command still gives the user the option of not enforcing this restriction by specifying the \st{dont\_drop\_larger\_lower} option. For all
$g$ such that $F_g\leq T_g$, let
\[
	S_g= \begin{cases} 1 \text{ if } D_{g,F_g}>D_{g,1} \\ 0 \text{ if } D_{g,F_g}<D_{g,1} \\ \text{ . undefined for never switchers }\end{cases}.
\]
When Design Restriction 2 in the paper is enforced, $S_g=1$ above just corresponds to the switchers-in in the paper, while $S_g=0$ corresponds to the switchers-out therein (therein, switchers-out are indexed by $S_g=-1$ instead of 0).  
\textit{Code tag: see ``Defining S\_g'' section in the code.\footnote{In the code, the definition of $S_g$ is slightly different from that given above, but the two definitions are equivalent when the \st{dont\_drop\_larger\_lower} option is not specified.}}

\paragraph{Event-study and average-total-effect estimators for switchers-in.}
Let $$L_u=\max_{g:S_g=1}(T_g-F_g+1)$$ denote the largest $\ell$ such that $\delta_{g,\ell}$ can be estimated, for groups $g$ such that $S_g=1$. For every $\ell \in \{1,...,L_u\}$, let
$$N_\ell^1=\sum_{g:S_g=1,F_g-1+\ell\leq T_g} N_{g,F_g-1+\ell}>0$$ be the number of units in groups reaching $\ell$ periods after their first treatment change at or before $T_g$, and such that $S_g=1$.  For every $\ell \in \{1,...,L_u\}$, let
\begin{equation}
\DID_{+,\ell} = \sum_{g:S_g=1,F_g-1+\ell\leq T_g} \frac{N_{g,F_g-1+\ell}}{N_\ell^1}\DID_{g,\ell}.	
	\label{eq:def_did_ellplus_nonbin}
\end{equation}
Let $w_{+,\ell}=N^1_\ell/\sum_{\ell'=1}^{L_u} N^1_{\ell'}$ (\textit{Code tag: see ``Computing the weights w\_+,l'' section in the code}), let
\begin{align*}
    \delta^D_{+,\ell}=&\sum_{g:S_g=1,F_g-1+\ell\leq T_g} \frac{N_{g,F_g-1+\ell}}{N_\ell^1}(D_{g,F_g-1+\ell}-D_{g,1}),
\end{align*}
(\textit{Code tag: see ``Computing the delta\_D\_+,l'' section in the code})
and let 
\begin{align*}
	 \widehat{\delta}_+& = \frac{\sum_{\ell=1}^{L_u} w_{+,\ell} \DID_{+,\ell}}{\sum_{\ell=1}^{L_u} w_{+,\ell} \delta^D_{+,\ell}}.
 \end{align*}

\paragraph{Event-study and average-total-effect for switchers-out.}
Let $$L_a=\max_{g:S_g=0}(T_g-F_g+1).$$ For every $\ell \in \{1,...,L_a\}$, let $$N_\ell^0=\sum_{g:S_g=0,F_g-1+\ell\leq T_g} N_{g,F_g-1+\ell}.$$  
For every $\ell \in \{1,...,L_a\}$, let
\begin{equation}
\DID_{-,\ell} = \sum_{g:S_g=0,F_g-1+\ell\leq T_g} \frac{N_{g,F_g-1+\ell}}{N_\ell^0}(-\DID_{g,\ell}).	
	\label{eq:def_did_ellminus_nonbin}
\end{equation}
Let $w_{-,\ell}=N^0_\ell/\sum_{\ell'=1}^{L_a} N^0_{\ell'}$, let
\begin{align*}
\delta^D_{-,\ell}=&\sum_{g:S_g=0,F_g-1+\ell\leq T_g} \frac{N_{g,F_g-1+\ell}}{N_\ell^0}(D_{g,1}-D_{g,F_g-1+\ell}),
\end{align*}
and let 
\begin{align*}
	 \widehat{\delta}_-& = \frac{\sum_{\ell=1}^{L_a} w_{-,\ell} \DID_{-,\ell}}{\sum_{\ell=1}^{L_a} w_{-,\ell} \delta^D_{-,\ell}}.
 \end{align*}

\paragraph{Aggregated event-study and average-total-effect across switchers-in and switchers-out.}
Let
\begin{equation}\label{eq:def_did_ell_sup}
\DID_{\ell} = \frac{N^1_\ell}{N^1_\ell+N^0_\ell}\DID_{+,\ell}+\frac{N^0_\ell}{N^1_\ell+N^0_\ell}\DID_{-,\ell}.
\end{equation}
By default, the event-study effects reported by \st{did\_multiplegt\_dyn} are the estimators $\DID_{\ell}$. Let $N_\ell=N^1_\ell+N^0_\ell$.
\cite{de2022intertemporal} show that $\DID_{\ell}$ is unbiased for the non-normalized event-study effect $$\delta_{\ell}:=\sum_{g:F_g-1+\ell\leq T_g} \frac{N_{g,F_g-1+\ell}}{N_\ell}\left(1\{S_g=1\}-1\{S_g=0\}\right)\delta_{g,\ell}.$$ $\delta_{g,\ell}$ is the effect, for switcher $g$, of having received their actual rather than their period-one treatment dose, for $\ell$ periods. Switchers-in' actual treatments are weakly larger than their period-one treatment, while switchers-out actual treatments are weakly lower. Therefore, for all $g$ $\left(1\{S_g=1\}-1\{S_g=0\}\right)\delta_{g,\ell}$ is an effect of having been exposed to a weakly larger dose for $\ell$ periods. $\delta_{\ell}$ averages those effects across all switchers. If the \st{effects\_equal} option is specified, the command performs an F-test that all estimated effects are equal. While non-normalized event-study effects can be interpreted as average effects of being exposed to a weakly higher treatment dose for $\ell$ periods, the magnitude and timing
    of the incremental treatment doses can vary across groups. The command's \st{design} option can be used to report switchers' period-one and subsequent treatments, thus helping the analyst understand the treatment paths whose effect is aggregated in the non-normalized event-study effects. The command's \st{by\_path} option can be used to compute treatment-path specific event-study effects.

\medskip
Let
\begin{align*}
    &w_+\\
    =& \frac{\sum_{g:S_g=1,F_g\leq T_g}\sum_{\ell=1}^{T_g-F_g+1}N_{g,F_g-1+\ell}(D_{g,F_g-1+\ell}-D_{g,1})}{\sum_{g:S_g=1,F_g\leq T_g}\sum_{\ell=1}^{T_g-F_g+1}N_{g,F_g-1+\ell}(D_{g,F_g-1+\ell}-D_{g,1})+\sum_{g:S_g=0,F_g\leq T_g}\sum_{\ell=1}^{T_g-F_g+1}N_{g,F_g-1+\ell}(D_{g,1}-D_{g,F_g-1+\ell})} \\
=& \frac{\sum_{\ell=1}^{L_u} w_{+,\ell} \delta^D_{+,\ell} \times (\sum_{\ell'=1}^{L_u} N^1_{\ell'}) }{\sum_{\ell=1}^{L_u} w_{+,\ell} \delta^D_{+,\ell} \times (\sum_{\ell'=1}^{L_u} N^1_{\ell'}) + \sum_{\ell=1}^{L_a} w_{-,\ell} \delta^D_{-,\ell} \times (\sum_{\ell'=1}^{L_a} N^0_{\ell'})}
\end{align*}
(\textit{Code tag: see ``Computing the weight w\_+'' section in the code}) and
\begin{equation}\label{eq:def_widehat_delta_sup}
\widehat{\delta}=w_+\widehat{\delta}_{+}+(1-w_+)\widehat{\delta}_{-}.
\end{equation}
\cite{de2022intertemporal} show that $\widehat{\delta}$ is unbiased for the average total effect per unit of treatment $\delta$ defined therein.

\paragraph{Normalized event-study estimators with the \st{normalized} option.}
For any $g$ such that $F_g\leq T_g$ and any $\ell \in \{1,...,T_g-F_g+1\}$, let
$$\delta^D_{g,\ell}=\sum_{k=0}^{\ell-1} (D_{g,F_g+k}-D_{g,1})$$
be the difference between the total treatment dose received by group $g$ from $F_g$ to $F_g-1+\ell$, and the total treatment dose it would have received in the status-quo counterfactual. Let $\delta^D_{\ell}=\sum_{g:F_{g}-1+\ell\leq T_{g}}\frac{N_{g,F_g-1+\ell}}{N^1_\ell+N^0_\ell}|\delta^D_{g,\ell}|$ (\textit{Code tag: see ``Computing aggregated delta\_D'' section in the code}), and let
\begin{equation}\label{eq:def_did_ell_sup}
\DID^n_{\ell} = \frac{\DID_{\ell}}{\delta^D_{\ell}}.
\end{equation}
(\textit{Code tag: see ``Computing DID\_n\_l'' section in the code})
\cite{de2022intertemporal} show that $\DID^n_{\ell}$ is unbiased for the normalized event-study effect $\delta^n_{\ell}$ defined therein, which measures a weighted average of the effects of the contemporaneous treatment and of its $\ell-1$ first lags on the contemporaneous outcome, with weights that can be estimated.
If the \st{normalized} option is specified, the event-study effects reported by \st{did\_multiplegt\_dyn} are the estimators $\DID^n_{\ell}$. If the \st{normalized\_weights} option is also specified, the command reports the weights that normalized effect $\ell$ puts on the effect of the current treatment, on the effect of the first treatment lag, etc. 

\subsection{Estimators' asymptotic distributions and variances.}  
\label{subsec:asymptotics}

\paragraph{Assumption to derive estimators' asymptotic distributions.}
Asymptotic results in the paper rely on the following assumption. 
\begin{hyp}
\label{hyp:indep_groups}
(Independent groups) Conditional on $(\bm{D}_g)_{g\ge 1}$, the vectors $(\bm{Y}_g)_{g\geq 1}$ are mutually independent.
\end{hyp}
Assumption \ref{hyp:indep_groups} allows for serial correlation of the treatments and outcomes within each group, which may be important features to account for in DID studies \citep{bertrand2004}. 

\paragraph{Writing the event-study and average-total-effect estimators for switchers-in as averages of group-level variables.}
For every $\ell \in \{1,...,L_u\}$, $t\in \{\ell+1,...,T_g\}$, and $g$, let
$$N^1_{t,\ell,g}=\sum_{g':S_{g'}=1, F_{g'}-1+\ell=t,D_{g',1}=D_{g,1}, N^{g'}_t>0} N_{g',t}.$$ 
Let
\begin{align*}
	U^+_{g,\ell} & = 1\{\ell\leq T_g-1\}\frac{G}{N^1_{\ell}} \sum_{t=\ell+1}^{T_g}  \left[\ind{F_g-1+\ell=t,S_g=1}- \frac{N^1_{t,\ell,g}}{N^g_t}\ind{F_g>t}\right] N_{g,t} (Y_{g,t}-Y_{g,t-\ell}),\\
	U^+_{g} & = \frac{\sum_{\ell=1}^{L_u} w_{+,\ell} U^+_{g,\ell}}{\sum_{\ell=1}^{L_u} w_{+,\ell} \delta^D_{+,\ell}}.
\end{align*}
(\textit{Code tag: see ``Computing U\_+\_(G,g,l)'' and ``Computing the U\_+\_(G,g)s'' sections in the code}).
One has
\begin{equation}\label{eq:lin7}
\DID_{+,\ell} = \frac{1}{G} \sum_{g=1}^G U^+_{g,\ell}
\end{equation}
and
\begin{equation}\label{eq:lin8}
\widehat{\delta}_+= \frac{1}{G}\sum_{g=1}^G U^+_{g}.
\end{equation}

\paragraph{Writing the event-study and average-total-effect estimators for switchers-out as averages of group-level variables.}
For every $\ell \in \{1,...,L_a\}$, $t\in \{\ell+1,...,T_g\}$, and $g$, let
$$N^0_{t,\ell,g}=\sum_{g':S_{g'}=0, F_{g'}-1+\ell=t,D_{g',1}=D_{g,1}, N^{g'}_t>0} N_{g',t}.$$ 
Let
\begin{align*}
	U^-_{g,\ell} & = 1\{\ell\leq T_g-1\}\frac{G}{N^0_{\ell}} \sum_{t=\ell+1}^{T_g} \left[\frac{N^0_{t,\ell,g}}{N^g_t}\ind{F_g>t}-\ind{F_g-1+\ell=t,S_g=0}\right]  N_{g,t} (Y_{g,t}-Y_{g,t-\ell}),\\
	U^-_{g} & = \frac{\sum_{\ell=1}^{L_a} w_{-,\ell} U^-_{g,\ell}}{\sum_{\ell=1}^{L_a} w_{-,\ell} \delta^D_{-,\ell}}.
\end{align*}
One has
\begin{equation}\label{eq:lin9}
\DID_{-,\ell} = \frac{1}{G} \sum_{g=1}^G U^-_{g,\ell}
\end{equation}
and
\begin{equation}\label{eq:lin10}
\widehat{\delta}_-= \frac{1}{G}\sum_{g=1}^G U^-_{g}.
\end{equation}

\paragraph{Writing the aggregated event-study and average-total-effect estimators as averages of group-level variables.}
Let
\begin{align}
U_{g,\ell}&=  \frac{N^1_\ell}{N^1_\ell+N^0_\ell} U^+_{g,\ell}+  \frac{N^0_\ell}{N^1_\ell+N^0_\ell} U^-_{g,\ell}\label{eq:fi5}\\
U^n_{g,\ell}&=  \frac{N^1_\ell}{N^1_\ell+N^0_\ell} \frac{U^+_{g,\ell}}{\delta^D_{\ell}}+  \frac{N^0_\ell}{N^1_\ell+N^0_\ell} \frac{U^-_{g,\ell}}{\delta^D_{\ell}}\label{eq:fi5'}\\
U_{g}&= w_+ U^+_{g}+ (1-w_+) U^-_{g}.\label{eq:fi6}
\end{align}
(\textit{Code tag: see the ``Aggregating the U\_(g,l) for switchers in and out'' section in the code}). 
Equations \eqref{eq:lin7}, \eqref{eq:lin8}, \eqref{eq:lin9}, \eqref{eq:lin10}, \eqref{eq:def_did_ell_sup}, and \eqref{eq:def_widehat_delta_sup} imply
that
\begin{equation}\label{eq:lin11}
\DID_{\ell} = \frac{1}{G} \sum_{g=1}^G U_{g,\ell},
\end{equation}
\begin{equation}\label{eq:lin11'}
\DID^n_{\ell} = \frac{1}{G} \sum_{g=1}^G U^n_{g,\ell},
\end{equation}
and
\begin{equation}\label{eq:lin12}
\widehat{\delta}= \frac{1}{G} \sum_{g=1}^GU_{g}.
\end{equation}
Conditional on the treatments, the only random quantities in $U_{g,\ell}$, $U^n_{g,\ell}$, and $U_{g}$ are $(Y_{g,1},...,Y_{g,T})$. Under Assumption \ref{hyp:indep_groups}, the vectors $(Y_{g,1},...,Y_{g,T})$ are independent across groups, so the 
variables $U_{g,\ell}$, $U^n_{g,\ell}$, and $U_{g}$ are also independent across groups. Thus \eqref{eq:lin11} (resp. \eqref{eq:lin11'}, \eqref{eq:lin12}) shows that $\DID_{\ell}$ (resp. $\DID^n_{\ell}$, $\widehat{\delta}$) is an average, across groups, of independent (though non identically distributed) variables. Then, consistency and asymptotic normality of those estimators follow from the weak law of large numbers and central limit theorem for independent but non identically distributed variables. 

\paragraph{Variance estimators.}
We consider an estimator of $\sigma^2_\ell=V(\DID_{\ell}|\bm{D})$ of the form
$$\widehat{\sigma}^2_\ell(\widehat{\theta}_1,...,\widehat{\theta}_G)=\frac{1}{G^2}\sum_{g=1}^G \left(U_{g,\ell} - \widehat{\theta}_g\right)^2,$$
where $\widehat{\theta}_g$ is an estimator of $E(U_{g,\ell}|\bm{D})$. In our set-up where we condition on $\bm{D}$, the $U_{g,\ell}$s are i.n.i.d, so in general 
we cannot unbiasedly estimate $E(U_{g,\ell}|\bm{D})$, as this expectation varies across groups. Then, letting 
$\theta_g$ denote the expectation of $\widehat{\theta}_g$ (assuming it exists), 
$\widehat{\sigma}^2_\ell(\widehat{\theta}_1,...,\widehat{\theta}_G)$ is asymptotically equivalent to
$$\widehat{\sigma}^2_\ell(\theta_1,...,\theta_G)=\frac{1}{G^2}\sum_{g=1}^G \left(U_{g,\ell} - \theta_g\right)^2.$$
Because $E\left(\left(U_{g,\ell} - \theta_g\right)^2\right)\geq V(U_{g,\ell})$ for any real number $\theta_g$, inference is in general conservative. 
However, as is often the case in i.n.i.d. set up, we can propose estimators $(\widehat{\theta}_g)_{g\ge 1}$ that lead to non-conservative inference, if the treatment effect is constant. In our dynamic potential outcome setup, we express this condition as follows. Let $s_{g,\ell,k}$ be the slope of $g$'s potential outcome function at $F_g-1+\ell$ with respect to its $k$th treatment lag defined in the paper. We assume that
\begin{equation}\label{eq:constant_effects}
s_{g,\ell,k}=s_{k},    
\end{equation}
meaning that effects of treatment lags are constant across groups and over time. Then,
\begin{align*}
E(U_{g,\ell}|\bm{D})=&  \frac{N^1_\ell}{N^1_\ell+N^0_\ell} E(U^+_{g,\ell}|\bm{D})+  \frac{N^0_\ell}{N^1_\ell+N^0_\ell} E(U^-_{g,\ell}|\bm{D}),\\
E(U^+_{g,\ell}|\bm{D})=&1\{\ell\leq T_g-1\}\frac{G}{N^1_{\ell}} \sum_{t=\ell+1}^{T_g}  \left[\ind{F_g-1+\ell=t,S_g=1}- \frac{N^1_{t,\ell,g}}{N^g_t}\ind{F_g>t}\right] N_{g,t}E((Y_{g,t}-Y_{g,t-\ell})|\bm{D}),\\
E(U^-_{g,\ell}|\bm{D})=&1\{\ell\leq T_g-1\}\frac{G}{N^0_{\ell}} \sum_{t=\ell+1}^{T_g} \left[\frac{N^0_{t,\ell,g}}{N^g_t}\ind{F_g>t}-\ind{F_g-1+\ell=t,S_g=0}\right] N_{g,t}E((Y_{g,t}-Y_{g,t-\ell})|\bm{D}).
\end{align*}
Therefore, if we can consistently estimate $E((Y_{g,t}-Y_{g,t-\ell})|\bm{D})$ for all $t<F_g$ and $t=F_g-1+\ell$, we can consistently estimate $E(U_{g,\ell}|\bm{D})$.
For $t<F_g$,
\begin{align}\label{eq:notgspecific1}
&E((Y_{g,t}-Y_{g,t-\ell})|\bm{D})\nonumber\\
=&E((Y_{g,t}(\bm{D}_{g,1,t})-Y_{g,t-\ell}(\bm{D}_{g,1,t-\ell}))|\bm{D})\nonumber\\
=&\phi(t,\ell,D_{g,1}),
\end{align}
for some function $\phi$, where the second equality follows from Assumption 2.
For $t=F_g-1+\ell$,
\begin{align}\label{eq:notgspecific2}
&E((Y_{g,F_g-1+\ell}-Y_{g,F_g-1})|\bm{D})\nonumber\\
=&E((Y_{g,F_g-1+\ell}(\bm{D}_{g,1,F_g-1+\ell})-Y_{g,F_g-1}(\bm{D}_{g,1,F_g-1}))|\bm{D})+\delta_{g,\ell}\nonumber\\
=&\phi(F_g-1+\ell,\ell,D_{g,1})+\sum_{k=0
}^{\ell-1}(D_{g,F_g-1+\ell-k}-D_{g,1})s_{k}\nonumber\\
=&\psi(F_g-1+\ell,\ell,D_{g,1},D_{g,F_g},...,D_{g,F_g-1+\ell}),
\end{align}
letting $$\psi(F_g-1+\ell,\ell,D_{g,1},D_{g,F_g},...,D_{g,F_g-1+\ell})=\phi(F_g-1+\ell,\ell,D_{g,1})+\sum_{k=0
}^{\ell-1}(D_{g,F_g-1+\ell-k}-D_{g,1})s_{k}.$$
The second equality follows from Assumption 2, Lemma 2 in the paper, and \eqref{eq:constant_effects}. 
The right hand side of \eqref{eq:notgspecific1} only depends on $g$ through $D_{g,1}$. The right hand side of  \eqref{eq:notgspecific2} only depends on $g$ through $(D_{g,1},F_g,D_{g,F_g},...,D_{g,F_g-1+\ell})$.
To simplify, let us assume for now that groups' treatment can change at most once:
\begin{equation}\label{eq:onetreatmentchange}
D_{g,t}=D_{g,F_g} \text{ for all }t\geq F_g.    
\end{equation}
Then, the right hand side of \eqref{eq:notgspecific2} only depends on $g$ through $(D_{g,1},F_g,D_{g,F_g})$.
For groups whose treatment never changes, let $D_{g,F_g}=D_{g,1}$. 

\medskip
Thus, in view of \eqref{eq:notgspecific1}, for $t<F_g$ we estimate $E((Y_{g,t}-Y_{g,t-\ell})|\bm{D})$ by the average of $Y_{g',t}-Y_{g',t-\ell}$ across all not-yet-switchers with the same baseline treatment as $g$ (including $g$ itself), provided there is at least one other such not-yet-switcher. Similarly, in view of \eqref{eq:notgspecific2}, for $t=F_g-1+\ell$, we estimate $E((Y_{g,F_g-1+\ell}-Y_{g,F_g-1})|\bm{D})$ by the average of $Y_{g',F_g-1+\ell}-Y_{g',F_g-1}$ across all switchers with the same ($D_{g',1},F_{g'},D_{g',F_{g'}})$, provided there is at least one other such switcher. When a not-yet-switcher cannot be matched to another not-yet-switcher with the same baseline treatment, we estimate $E((Y_{g,t}-Y_{g,t-\ell})|\bm{D})$ by the average of $Y_{g',t}-Y_{g',t-\ell}$ across all not-yet-switchers and switchers such that $t=F_g-1+\ell$ and with the same baseline treatment as $g$. Similarly, if a switcher cannot be matched with another switcher with the same ($D_{g,1},F_{g},D_{g,F_{g}})$, we estimate $E((Y_{g,F_g-1+\ell}-Y_{g,F_g-1})|\bm{D})$ by the average of $Y_{g,F_g-1+\ell}-Y_{g,F_g-1}$ across all switchers with the same ($D_{g,1},F_{g}$) and not yet switchers with the same $D_{g,1}.$ The variance estimator with the correspondings $\widehat{\theta}_g$ is valid if the treatment effect is constant, if groups' treatment can only change once, and if not-yet-switchers can always be matched with another not-yet-switcher with the same $D_{g,1}$ and switchers can always be matched with another switcher with the same ($D_{g,1},F_{g},D_{g,F_{g}})$. Otherwise, our variance estimators are conservative. In particular, when a switcher cannot be matched with another switcher with the same ($D_{g,1},F_{g},D_{g,F_{g}})$, matching them with not-yet-switchers to estimate $E((Y_{g,F_g-1+\ell}-Y_{g,F_g-1})|\bm{D})$ will yield a biased estimator, and therefore a conservative variance estimator, if the treatment has an effect and therefore $E((Y_{g,t}-Y_{g,t-\ell})|\bm{D})$ differs between switchers and not-yet-switchers.

\medskip
Formally, let $\mathcal{G} = \{g: D_{g,1} \in \mathcal{D}_1^r\}$ be the set of groups $g$ such that there exists another group $g'$ with the baseline treatment $D_{g',1} = D_{g,1}$ and first-switch period $F_{g'} \neq F_g$. Let $f$ be defined as
$$
f(g) = (D_{g,1}, F_g, D_{g,F_g}) 
\ \ \ \ \ 
(g \in \mathcal{G})
$$
that is, $f$ maps each group $g$ from $\mathcal{G}$ to a triple consisting of group $g$'s status-quo treatment, first-switch period and treatment at switch. In the spirit of Assumption 5 in the paper, we assume that the number of distinct values of $(D_{g,1},F_g,D_{g,F_g})_{g \in \mathcal{G}}$, i.e. the image of $f$ in $\mathcal{G}$, is finite. Let
$$
\mathcal{C}_g = f^{-1}(f(g)) = \{g' \in \mathcal{G}: f(g') =  (D_{g,1}, F_g, D_{g,F_g})\}
$$
denote the set of groups in $\mathcal{G}$ that share the same status-quo treatment, first-switch period and treatment at switch as group $g$. Hereafter, we denote $\mathcal{C}_g$ as the \emph{cohort} of $g$. Similarly, let
$$
\mathcal{N}_{d,t} = \bigcup_{g: D_{g,1} = d, F_g > t} \mathcal{C}_g 
$$
be the set of not-yet-switchers at time $t$ with baseline treatment $d$, and
$$
\mathcal{S}^+_{d,t, \ell} = \bigcup_{\substack{g: D_{g,1} = d, F_g = t - \ell + 1, \\ S_g = 1}} \mathcal{C}_g 
\ \ \ \ \ \ \ \ 
\mathcal{S}^-_{d,t, \ell} = \bigcup_{\substack{g: D_{g,1} = d, F_g = t - \ell + 1, \\ S_g = 0}} \mathcal{C}_g 
$$
be the set of switchers-in and switchers-out that at time $t$ are $\ell$ periods away from their first switch (henceforth, $\ell$-switchers-in and $\ell$-switchers-out, respectively), and that have baseline treatment $d$. Lastly, let
$$
\mathcal{U}^+_{d,t, \ell} = \mathcal{N}_{d,t} \cup \mathcal{S}^+_{d,t,\ell}
\ \ \ \ \ \ \ \ 
\mathcal{U}^-_{d,t, \ell} = \mathcal{N}_{d,t} \cup \mathcal{S}^-_{d,t,\ell}
$$
denote the set of $\ell$-switchers-in and not-yet-switchers at $t$, and $\ell$-switchers-out and not-yet-switchers at $t$, with baseline treatment $d$. Notice that, while it is possible that $\#\mathcal{C}_g = 1$ or $\#\mathcal{N}_{d,t} = 1$, it holds that $\#\mathcal{U}^+_{d,t,\ell} > 1$ or $\#\mathcal{U}^-_{d,t,\ell} > 1$ for all $d \in \mathcal{D}_1^r$ and $t \in \{1, ..., T_g\}, \ell \in \{1, ..., T_g - F_g + 1\}$ for some $g \in \mathcal{G}$. More specifically, in cases where there are both switchers-in and switchers-out, both conditions hold. Conversely, in cases with only switchers-in (switchers-out), only the first (second) condition is guaranteed to hold. As a result, we can define
$$
\mathcal{B}^{S,+}_{g,t, \ell} =
\begin{cases}
\mathcal{C}_g & \text{ if } \#\mathcal{C}_g > 1 \\
\mathcal{U}^+_{D_{g,1}, t, \ell} & \text{ if } \#\mathcal{C}_g = 1
\end{cases}
\hspace{2cm}
\mathcal{B}^{N,+}_{g,t, \ell} =
\begin{cases}
\mathcal{N}_{D_{g,1}, t} & \text{ if } \#\mathcal{N}_{D_{g,1}, t} > 1 \\
\mathcal{U}^+_{D_{g,1}, t, \ell} & \text{ if } \#\mathcal{N}_{D_{g,1},t} = 1
\end{cases}
$$
$$
\mathcal{B}^{S,-}_{g,t, \ell} =
\begin{cases}
\mathcal{C}_g & \text{ if } \#\mathcal{C}_g > 1 \\
\mathcal{U}^-_{D_{g,1}, t, \ell} & \text{ if } \#\mathcal{C}_g = 1
\end{cases}
\hspace{2cm}
\mathcal{B}^{N,-}_{g,t, \ell} =
\begin{cases}
\mathcal{N}_{D_{g,1}, t} & \text{ if } \#\mathcal{N}_{D_{g,1}, t} > 1 \\
\mathcal{U}^-_{D_{g,1}, t, \ell} & \text{ if } \#\mathcal{N}_{D_{g,1},t} = 1
\end{cases}
$$

Let
\[
\widehat{E}^+_{g,t,\ell}=
\begin{cases}
\frac{\sum_{g'\in \mathcal{B}^{N,+}_{g,t, \ell}} N_{g',t}(Y_{g',t}-Y_{g',t-\ell})}{\sum_{g'\in\mathcal{B}^{N,+}_{g,t, \ell}}N_{g',t}} & \text{ if } t < F_g \\ 
\noalign{\vskip9pt}
\frac{\sum_{g'\in \mathcal{B}^{S,+}_{g,t, \ell}} N_{g',t}(Y_{g',t}-Y_{g',t-\ell})}{\sum_{g'\in\mathcal{B}^{S,+}_{g,t, \ell}}N_{g',t}} & \text{ if } t = F_g - 1 + \ell, S_g = 1 \\
\noalign{\vskip9pt}
\cdot & \text{ otherwise }
\end{cases}
\]
\[
\widehat{E}^-_{g,t,\ell}=
\begin{cases}
\frac{\sum_{g'\in \mathcal{B}^{N,-}_{g,t, \ell}} N_{g',t}(Y_{g',t}-Y_{g',t-\ell})}{\sum_{g'\in\mathcal{B}^{N,-}_{g,t, \ell}}N_{g',t}} & \text{ if } t < F_g \\ 
\noalign{\vskip9pt}
\frac{\sum_{g'\in \mathcal{B}^{S,-}_{g,t, \ell}} N_{g',t}(Y_{g',t}-Y_{g',t-\ell})}{\sum_{g'\in\mathcal{B}^{S,-}_{g,t, \ell}}N_{g',t}} & \text{ if } t = F_g - 1 + \ell, S_g = 0 \\
\noalign{\vskip9pt}
\cdot & \text{ otherwise }
\end{cases}
\]
\textit{Code tag: see ``E\_hat\_(g,t)'' section in the code.} 
Let
$$
\text{DOF}^+_{g,t, \ell}=\ind{F_g-1+\ell=t, S_g = 1}\sqrt{\frac{\#\mathcal{B}^{S,+}_{g, t, \ell}}{\#\mathcal{B}^{S,+}_{g,t,\ell}-1}}+ \ind{t<F_g}\sqrt{\frac{\#\mathcal{B}^{N,+}_{g,t, \ell}}{\#\mathcal{B}^{N,+}_{g,t, \ell}-1}}
$$
$$
\text{DOF}^-_{g,t, \ell}=\ind{F_g-1+\ell=t, S_g = 0}\sqrt{\frac{\#\mathcal{B}^{S,-}_{g, t, \ell}}{\#\mathcal{B}^{S,-}_{g,t,\ell}-1}}+ \ind{t<F_g}\sqrt{\frac{\#\mathcal{B}^{N,-}_{g,t, \ell}}{\#\mathcal{B}^{N,-}_{g,t, \ell}-1}}
$$

\textit{Code tag: see ``DOF\_(g,t)'' section in the code.}

Finally, let

\small
\begin{align*}
U^{+,var}_{g,\ell} &= 1\{\ell\leq T_g-1\}\frac{G}{N^1_{\ell}} \sum_{t=\ell+1}^{T_g}\left[\ind{F_g-1+\ell=t,S_g=1}- \frac{N^1_{t,\ell,g}}{N^g_t}\ind{F_g>t}\right] N_{g,t}\text{DOF}^+_{g,t,\ell}\left(Y_{g,t}-Y_{g,t-\ell}-\widehat{E}^+_{g,t,\ell}\right)\\
U^{-,var}_{g,\ell}& =1\{\ell\leq T_g-1\}\frac{G}{N^0_{\ell}} \sum_{t=\ell+1}^{T_g} \left[\frac{N^0_{t,\ell,g}}{N^g_t}\ind{F_g>t}-\ind{F_g-1+\ell=t,S_g=0}\right]N_{g,t}\text{DOF}^-_{g,t,\ell}\left(Y_{g,t}-Y_{g,t-\ell}-\widehat{E}^-_{g,t,\ell}\right)\\
U^{+,var}_{g} & = \frac{\sum_{\ell=1}^{L_u} w_{+,\ell} U^{+,var}_{g,\ell}}{\sum_{\ell=1}^{L_u} w_{+,\ell} \delta^D_{+,\ell}}\\
	U^{-,var}_{g} & = \frac{\sum_{\ell=1}^{L_a} w_{-,\ell} U^{-,var}_{g,\ell}}{\sum_{\ell=1}^{L_a} w_{-,\ell} \delta^D_{-,\ell}}\\
U^{var}_{g,\ell}&=  \frac{N^1_\ell}{N^1_\ell+N^0_\ell} U^{+,var}_{g,\ell}+  \frac{N^0_\ell}{N^1_\ell+N^0_\ell} U^{-,var}_{g,\ell}\\
U^{var}_{g}&= w_+ U^{+,var}_{g}+ (1-w_+) U^{-,var}_{g}.
\end{align*}
\normalsize
\textit{Code tag: see the ``Computing U\_(+,var)\_(G,g,l)'' section in the code.}
Finally, we define the variance estimators of $\DID_\ell$, $\DID^n_\ell$, and $\delta$ as
\begin{align}
\widehat{\sigma}^2_\ell=&\frac{1}{G^2}\sum_{g=1}^G \left(U^{var}_{g,\ell}\right)^2, \label{eq:var1}\\
\widehat{\sigma}^2_{\ell,n}=&\frac{\widehat{\sigma}^2_\ell}{(\delta^D_{\ell})^2},\label{eq:var1'}\\
\widehat{\sigma}^2=&\frac{1}{G^2}\sum_{g=1}^G \left(U^{var}_{g}\right)^2.\label{eq:var2}
\end{align}
\textit{Code tag: see the ``Compute sigma\_hat\_2\_l'' section in the code.}

\paragraph{Less-conservative variance estimators when groups’ treatment can change more than once.} When the \st{more\_granular\_demeaning} option is specified, the command computes variance
estimators that might be less conservative than those implemented by default, when groups’ treatment can change more than once. Then the cohorts with respect to which the $Y_{g,t}-Y_{g,t-\ell}$ of switchers is demeaned are defined by $(D_{g,1},F_g,D_{g,F_g} , ...,D_{g,F_g-1+\ell})$. This may lead to less conservative standard errors, when cohorts defined by $(D_{g,1},F_g,D_{g,F_g} , ...,D_{g,F_g-1+\ell})$ all contain at least two switchers. 


\section{Extensions}

\subsection{Placebos: \st{placebo(\#)}}
\label{sec:placebos}

The command can compute placebo estimators  $\DID^{\text{pl}}_{\ell}$, to test the no-anticipation and parallel-trends assumptions underlying the event-study estimators it computes. Those placebos are defined as symmetrically as possible to the actual event-study estimators. Namely, $\DID^{\text{pl}}_{\ell}$ is defined exactly as $\DID_{\ell}$, replacing
$Y_{g,F_g-1+\ell} - Y_{g,F_g-1}$ by $Y_{g,F_g-1-\ell} - Y_{g,F_g-1}$. 
Intuitively, $\DID^{\text{pl}}_{\ell}$ compares the outcome evolution of switchers included in $\DID_{\ell}$ to that of their control groups, from period $F_g-\ell-1$ to period $F_g-1$, namely before switchers' treatment changes for the first time. Accordingly, $\DID^{\text{pl}}_{\ell}$ assesses if those switchers and their control groups experience parallel evolutions of their status-quo potential outcomes, for $\ell$ periods, the number of periods over which parallel trends has to hold for $\DID_{\ell}$ to be unbiased. There may be switchers included in $\DID_{\ell}$ such that $F_g< \ell+2$: those switchers have to be excluded from the placebo, because they are not observed for sufficiently many time periods prior to their first treatment change, and $Y_{g,F_g-1-\ell} - Y_{g,F_g-1}$ cannot be computed for them. With a balanced panel those are the only switchers included in  $\DID_{\ell}$ that have to be excluded from $\DID^{\text{pl}}_{\ell}$. Conversely, there may be switchers for which $Y_{g,F_g-1-\ell} - Y_{g,F_g-1}$ can be computed but $Y_{g,F_g-1+\ell} - Y_{g,F_g-1}$ cannot: those switchers are not included in $\DID_{\ell}$, so the command does not included them in  $\DID^{\text{pl}}_{\ell}$, so that the placebo compares switchers' and controls' outcome evolution only for switchers included in the event-study effect. This is also the reason why the number of placebos requested cannot be larger than the number of effects requested. Note that while $\DID_{\ell}$ goes from the past (period $F_g-1$) to the future (period $F_g-1+\ell$), $\DID^{\pl}_{\ell}$ goes from the future (period $F_g-1$) to the past (period $F_g-1-\ell$). This follows the standard practice in event-study regressions, where the reference period is the one before the event. See Section 1.1 of the paper's Web Appendix for further details. If the user requests that at least two placebos be estimated,
the command computes the p-value of a joint test that all placebos are equal. 

\subsection{Estimation with controls: \st{controls(varlist)}}  
\label{sec:controls}

\paragraph{Setup.}
In this section, we consider estimators with covariates. Covariates are conditioned upon so that in this section as well, only groups' potential outcomes are random. Let $X_{g,t}$ denote the vector of controls for group $g$ at time $t$, inputted into the \st{controls} option. Those variables have to be time-varying, but to control for time-invariant covariates $X_g$, one can just let $X_{g,t}=X_g\times t$, or one can let $X_{g,t}$ be a vector of interactions between $X_g$ and
$(1\{t\geq t'\})_{t'\in \{1,...,T\}}$, indicators for period $t$ being after $t'$. Estimators with controls are similar to the baseline estimators presented in the previous section, except that $Y_{g,t} - Y_{g,t-\ell}$ is replaced by $Y_{g,t} - Y_{g,t-\ell}- (X_{g,t} - X_{g,t-\ell})'\widehat{\theta}_{D_{g,1}}$ in their definition, where $\widehat{\theta}_{D_{g,1}}$ is estimated as explained below. In this companion note, we only present estimators of non-normalized event-study effects with controls, and estimators of the variances of those estimators. With those estimators, it is easy to construct estimators of normalized event-study effects, an estimator of the average total effect, as well as estimators of the variances of those estimators: the command computes estimators of normalized event-study effects with controls, an estimator of the average total effect with controls, and the variances of those estimators.

\paragraph{Identifying assumption.}
With controls, we replace Assumption \ref{hyp:strong_exogeneity} by the following parallel trends assumption: 
\begin{hyp}\label{hyp:conds_withX}
(Parallel trends with covariates) There are vectors $(\theta_d)_{d\in\mathcal{D}_1^{\text{r}}}$ of same dimension as $X_{g,t}$ such that $\forall (g,g')$ such that $D_{g,1}=D_{g',1}\in\mathcal{D}_1^{\text{r}}$, then $\forall t\geq 2$,
\begin{align*}
& E[Y_{g,t}(\bm{D}_{g,1,t}) - Y_{g,t-1}(\bm{D}_{g,1,t-1}) - (X_{g,t} - X_{g,t-1})'\theta_{D_{g,1}} | \bm{D},\bm{X}] \notag\\
= & E[Y_{g’,t}(\bm{D}_{g’,1,t}) - Y_{g’,t-1}(\bm{D}_{g’,1,t-1}) - (X_{g',t} - X_{g',t-1})'\theta_{D_{g',1}} | \bm{D},\bm{X}]. \end{align*}
\end{hyp}
Assumption \ref{hyp:conds_withX} is similar to Assumption \ref{hyp:strong_exogeneity}, except that now groups can experience differential trends, provided those differential trends are fully explained by changes in their covariates. Note that the effect of these covariates is allowed to vary with groups' period-one treatment. With time-invariant covariates, if $X_{g,t}=X_g\times t$ then $(X_{g,t} - X_{g,t-1})'\theta_{D_{g,1}}=X_{g}'\theta_{D_{g,1}}$, so Assumption \ref{hyp:conds_withX} amounts to a conditional parallel-trends assumption, with a linear time-invariant functional form for the conditional counterfactual trend. Alternatively, if $X_{g,t}$ is a vector of interactions between $X_g$ and
$(1\{t\geq t'\})_{t'\in \{1,...,T\}}$, then, Assumption \ref{hyp:conds_withX} amounts to a conditional parallel-trends assumption, with a linear period-specific functional form for the conditional counterfactual trend.

\paragraph{Writing the difference between the sample and population coefficients on the controls as averages of group-level variables.} Let $X_{g,t}  = (X^{1}_{g,t}, \dots, X^{K}_{g,t})'$ be a $K\times 1$ vector of all the covariates of group $g$ at time $t$, 
let $\Delta X_{g,t}=X_{g,t}-X_{g,t-1}$, and let $\Delta Y_{g,t}=Y_{g,t}-Y_{g,t-1}$. For any $d\in \mathcal{D}^{\text{r}}_1$, let $T^d=\max_{g:D_{g,1}=d}F_g-1$, and let  $\widehat{\theta}_d = (\widehat{\theta}^{1}_{g,t}, \dots, \widehat{\theta}^{K}_{g,t})'$ denote the $K\times 1$ vector of sample coefficients of $\Delta X_{g,t}$ in the OLS regression of $\Delta Y_{g,t}$ on $\Delta X_{g,t}$ and time fixed effects, in the subsample  $ \mathcal{C}^d :\equiv \big\{(g,t) : D_{g,1}=d,F_g>t,2\leq t\leq T^d\big\}$.  Let $\theta_d$ denote the population coefficient of this regression. For every $2\leq t\leq T^d$, let  $$\Delta X^d_{.,t}=\sum_{g:D_{g,1}=d,F_g>t}\frac{N_{g,t}}{\sum_{g:D_{g,1}=d,F_g>t}N_{g,t}}\Delta X_{g,t}$$ be the $K\times 1$ vector equal to the average of $\Delta X_{g,t}$ across all $g$ such that $(g,t)\in \mathcal{C}^d$. The vector of residuals from the regressions of each element of $\Delta X_{g,t}$ on time FEs is $\Delta X_{g,t}-\Delta X^d_{.,t}$. Let $\Delta \Dot{X}_{g,t}$ denote this vector of residuals. By the Frisch-Waugh theorem,
\begin{equation*}
\widehat{\theta}_d= \left( \frac{1}{N^c_d} \sum_{(g,t) \in \mathcal{C}^d} N_{g,t}\Delta \Dot{X}_{g,t} \Delta \Dot{X}_{g,t}' \right)^{-1} \times \frac{1}{N^c_d} \sum_{(g,t) \in \mathcal{C}^d} N_{g,t} \Delta \Dot{X}_{g,t} \Delta Y_{g,t},
\end{equation*}
where $N^c_d=\sum_{(g,t)\in \mathcal{C}^d}N_{g,t}$. 
Letting $Den_d=\frac{1}{N^c_d} \sum_{(g,t) \in \mathcal{C}^d} N_{g,t}\Delta \Dot{X}_{g,t} \Delta \Dot{X}_{g,t}'$, let
\begin{align*}
	V^{d}_{g} & = Den_d^{-1} \frac{G}{N^c_d}\left(\sum_{t=2}^{F_g-1}N_{g,t}\Delta \Dot{X}_{g,t} \Delta Y_{g,t}\right)1\{D_{g,1}=d,F_g\geq 3\}-\theta_d.
\end{align*}
Note that conditional on the treatments and covariates, $Den_d$ is deterministic, so the random variables in $V^{d}_{g}$ are specific to group $g$, so the variables $V^{d}_{g}$ are independent across $g$s.
 One has
\begin{equation}\label{eq:lin_OLS}
\widehat{\theta}_d-\theta_d=\frac{1}{G}\sum_{g=1}^G V^{d}_{g}.
\end{equation}
(\textit{Code tag: see the ``Necessary pre-estimation steps when the controls option is specified'' section in the code.})

\paragraph{Defining event-study estimators with controls.}
To estimate $\delta_{g,\ell}$, we use
\begin{align*}
\DID^X_{g,\ell} =&Y_{g,F_g-1+\ell} - Y_{g,F_g-1} -(X_{g,F_g-1+\ell} - X_{g,F_g-1})'\widehat{\theta}_{D_{g,1}}\\
-&\sum_{g':D_{g',1}=D_{g,1},F_{g'} >F_g-1+\ell} \frac{N_{g',F_g-1+\ell}}{N^{g}_{F_g-1+\ell}}(Y_{g',F_g-1+\ell} - Y_{g',F_g-1}-(X_{g',F_g-1+\ell} - X_{g',F_g-1})'\widehat{\theta}_{D_{g,1}}).
\end{align*}
For every $\ell \in \{1,...,L_u\}$, let
\begin{equation}
\DID^X_{+,\ell} = \sum_{g:S_g=1,F_g-1+\ell\leq T_g} \frac{N_{g,F_g-1+\ell}}{N_\ell^1}\DID^X_{g,\ell}.	
	\label{eq:def_did_ellplus_nonbinX}
\end{equation}
For every $\ell \in \{1,...,L_a\}$, let
\begin{equation}
\DID^X_{-,\ell} = \sum_{g:S_g=0,F_g-1+\ell\leq T_g} \frac{N_{g,F_g-1+\ell}}{N_\ell^0}(-\DID^X_{g,\ell}).	
	\label{eq:def_did_ellminus_nonbinX}
\end{equation}
Finally, let
\begin{equation}\label{eq:def_did_ell_x}
\DID^X_{\ell} = \frac{N^1_\ell}{N^1_\ell+N^0_\ell}\DID^X_{+,\ell}+\frac{N^0_\ell}{N^1_\ell+N^0_\ell}\DID^X_{-,\ell},
\end{equation}

\paragraph{Defining an infeasible event-study estimator with controls.}
Let
\begin{align*}
\widetilde{\DID}^X_{g,\ell} =&Y_{g,F_g-1+\ell} - Y_{g,F_g-1} -(X_{g,F_g-1+\ell} - X_{g,F_g-1})'\theta_{D_{g,1}}\\
-&\sum_{g':D_{g',1}=D_{g,1},F_{g'} >F_g-1+\ell} \frac{N_{g',F_g-1+\ell}}{N^{g}_{F_g-1+\ell}}(Y_{g',F_g-1+\ell} - Y_{g',F_g-1}-(X_{g',F_g-1+\ell} - X_{g',F_g-1})'\theta_{D_{g,1}}),
\end{align*}
\begin{equation*}
\widetilde{\DID}^X_{+,\ell} = \sum_{g:S_g=1,F_g-1+\ell\leq T_g} \frac{N_{g,F_g-1+\ell}}{N_\ell^1}\widetilde{\DID}^X_{g,\ell},
\end{equation*}
and
\begin{equation*}
\widetilde{\DID}^X_{-,\ell} = \sum_{g:S_g=0,F_g-1+\ell\leq T_g} \frac{N_{g,F_g-1+\ell}}{N_\ell^0}(-\widetilde{\DID}^X_{g,\ell}).	
\end{equation*}

\paragraph{A formula relating the feasible and infeasible event-study estimators with controls.}
Let
\begin{align*}
M^+_{d,\ell}=&\sum_{g:D_{g,1}=d,S_g=1,F_g-1+\ell\leq T_g} \frac{N_{g,F_g-1+\ell}}{N_\ell^1}\left[(X_{g,F_g-1+\ell} - X_{g,F_g-1})'\right.\nonumber\\
-&\left.\sum_{g':D_{g',1}=D_{g,1},F_{g'} >F_g-1+\ell} \frac{N_{g',F_g-1+\ell}}{N^{g}_{F_g-1+\ell}}(X_{g',F_g-1+\ell} - X_{g',F_g-1})'\right],
\end{align*}
\begin{align*}
M^-_{d,\ell}=&\sum_{g:D_{g,1}=d,S_g=0,F_g-1+\ell\leq T_g} \frac{N_{g,F_g-1+\ell}}{N_\ell^0}\left[(X_{g,F_g-1+\ell} - X_{g,F_g-1})'\right.\nonumber\\
-&\left.\sum_{g':D_{g',1}=D_{g,1},F_{g'} >F_g-1+\ell} \frac{N_{g',F_g-1+\ell}}{N^{g}_{F_g-1+\ell}}(X_{g',F_g-1+\ell} - X_{g',F_g-1})'\right],
\end{align*}
with the convention that a summation over an empty set is equal to 0. Note that conditional on the treatments and covariates, $M_{d,\ell}$ is deterministic. 
One has 
\begin{align}\label{eq:link_DIDX+knownunknowntheta}
&\DID^X_{+,\ell}=\widetilde{\DID}^X_{+,\ell}-\sum_{d\in \mathcal{D}^{\text{r}}_1}M^+_{d,\ell}(\widehat{\theta}_d-\theta_d)
\end{align}
\begin{align}\label{eq:link_DIDX-knownunknowntheta}
&\DID^X_{-,\ell}=\widetilde{\DID}^X_{-,\ell}+\sum_{d\in \mathcal{D}^{\text{r}}_1}M^-_{d,\ell}(\widehat{\theta}_d-\theta_d).
\end{align}

\paragraph{A decomposition of $M^+_{d,\ell}$ and $M^-_{d,\ell}$ as averages of group-level variables (only useful to compute $M^+_{d,\ell}$ and $M^-_{d,\ell}$, not to derive variances).}
One has 
\begin{equation}
    M^+_{d,\ell} = \frac{1}{G} \sum_{g=1}^G m^{+}_{g,d,\ell}
\end{equation}
and:
\begin{equation}
    M^-_{d,\ell} = \frac{1}{G} \sum_{g=1}^G m^{-}_{g,d,\ell}
\end{equation}
where: \\
$ m^{+}_{g,d,\ell} = 1\{\ell\leq T_g-2, D_{g,1} = d\}\frac{G}{N^1_{\ell}} \sum_{t=\ell+1}^{T_g}\left(\left[\ind{F_g-1+\ell=t,S_g=1}- \frac{N^1_{t,\ell,g}}{N^g_t}\ind{F_g>t}\right]\right.N_{g,t}\times(X_{g,t} - X_{g,t-\ell})'\Big) $ \\
and: \\
$ m^{-}_{g,d,\ell} = 1\{\ell\leq T_g-2, D_{g,1} = d\}\frac{G}{N^0_{\ell}} \sum_{t=\ell+1}^{T_g}\left( \left[\ind{F_g-1+\ell=t,S_g=0}- \frac{N^0_{t,\ell,g}}{N^g_t}\ind{F_g>t}\right]\right.\times N_{g,t}(X_{g,t} - X_{g,t-\ell})'\Big). $  \\[0.2cm]
(\textit{Code tag: see the ``Computing coordinates of vectors m\_+\_(g,d,l) and m\_-\_(g,d,l)'' section in the code.})

\paragraph{A decomposition of the infeasible event-study estimators with controls as averages of group-level variables.}
Let
\begin{align*}
	\widetilde{U}^{+,X}_{g,\ell} & = 1\{\ell\leq T_g-1\}\frac{G}{N^1_{\ell}} \sum_{t=\ell+1}^{T_g}\left(  \left[\ind{F_g-1+\ell=t,S_g=1}- \frac{N^1_{t,\ell,g}}{N^g_t}\ind{F_g>t}\right]\right.\\
&\left.\times N_{g,t}(Y_{g,t}-Y_{g,t-\ell}-(X_{g,t} - X_{g,t-\ell})'\theta_{D_{g,1}})\right).
\end{align*}
One has
\begin{equation}\label{eq:lin13}
\widetilde{\DID}^X_{+,\ell} = \frac{1}{G} \sum_{g=1}^G \widetilde{U}^{+,X}_{g,\ell}.
\end{equation}
Similarly, let
\begin{align*}
	\widetilde{U}^{-,X}_{g,\ell} & = 1\{\ell\leq T_g-1\}\frac{G}{N^0_{\ell}} \sum_{t=\ell+1}^{T_g}\left( N_{g,t} \left[\frac{N^0_{t,\ell,g}}{N^g_t}\ind{F_g>t} - \ind{F_g-1+\ell=t,S_g=0}\right]\right.\\
&\left.\times (Y_{g,t}-Y_{g,t-\ell}-(X_{g,t} - X_{g,t-\ell})'\theta_{D_{g,1}})\right).
\end{align*}
One has
\begin{equation}\label{eq:lin14}
\widetilde{\DID}^X_{-,\ell} = \frac{1}{G} \sum_{g=1}^G \widetilde{U}^{-,X}_{g,\ell}
\end{equation}

\paragraph{A decomposition of the feasible event-study estimators with controls as averages of group-level variables.}
Let
\begin{equation*}
U^{+,X}_{g,\ell}=\widetilde{U}^{+,X}_{g,\ell}-\sum_{d\in \mathcal{D}^{\text{r}}_1}M^+_{d,\ell}V^{d}_{g}.
\end{equation*}
It follows from Equations \eqref{eq:link_DIDX+knownunknowntheta}, \eqref{eq:lin_OLS}, and \eqref{eq:lin13} that
\begin{equation}\label{eq:lin15}
\DID^X_{+,\ell} = \frac{1}{G} \sum_{g=1}^G U^{+,X}_{g,\ell}.
\end{equation}
Similarly, let
\begin{equation*}
U^{-,X}_{g,\ell}=\widetilde{U}^{-,X}_{g,\ell}+\sum_{d\in \mathcal{D}^{\text{r}}_1}M^-_{d,\ell}V^{d}_{g}.
\end{equation*}
It follows from Equations \eqref{eq:link_DIDX-knownunknowntheta}, \eqref{eq:lin_OLS}, and \eqref{eq:lin14} that
\begin{equation}\label{eq:lin16}
\DID^X_{-,\ell} = \frac{1}{G} \sum_{g=1}^G U^{-,X}_{g,\ell}.
\end{equation}
Finally, let
\begin{align}
U^X_{g,\ell}&=  \frac{N^1_\ell}{N^1_\ell+N^0_\ell} U^{X,+}_{g,\ell}+  \frac{N^0_\ell}{N^1_\ell+N^0_\ell} U^{X,-}_{g,\ell}\label{eq:fi5}.
\end{align}
It follows from \eqref{eq:def_did_ell_x}, \eqref{eq:lin15}, and \eqref{eq:lin16} that
\begin{equation}
\DID^X_{\ell} = \frac{1}{G} \sum_{g=1}^G U^X_{g,\ell}.
\end{equation}

\paragraph{Variance estimator.}
Note that
\begin{align*}
E(U^{+,X}_{g,\ell}|\bm{D},\bm{X})& = 1\{\ell\leq T_g-1\}\frac{G}{N^1_{\ell}} \sum_{t=\ell+1}^{T_g}  \left[\ind{F_g-1+\ell=t,S_g=1}- \frac{N^1_{t,\ell,g}}{N^g_t}\ind{F_g>t}\right]\\
&\times N_{g,t} E((Y_{g,t}-Y_{g,t-\ell}-(X_{g,t} - X_{g,t-\ell})'\theta_{D_{g,1}})|\bm{D},\bm{X})\\
-&\sum_{d\in \mathcal{D}^{\text{r}}_1}M^+_{d,\ell}\left(Den_d^{-1} \frac{G}{N^c_d}\sum_{t=2}^{F_g-1}N_{g,t}\Delta \Dot{X}_{g,t} E\left(\Delta Y_{g,t}|\bm{D},\bm{X}\right)1\{D_{g,1}=d,F_g\geq 3\}-\theta_d\right)\\
E(U^{-,X}_{g,\ell}|\bm{D},\bm{X})& =1\{\ell\leq T_g-1\}\frac{G}{N^0_{\ell}} \sum_{t=\ell+1}^{T_g} \left[\frac{N^0_{t,\ell,g}}{N^g_t}\ind{F_g>t}-\ind{F_g-1+\ell=t,S_g=0}\right]\\
&\times N_{g,t}E((Y_{g,t}-Y_{g,t-\ell}-(X_{g,t} - X_{g,t-\ell})'\theta_{D_{g,1}})|\bm{D},\bm{X})\\
+&\sum_{d\in \mathcal{D}^{\text{r}}_1}M^-_{d,\ell}\left(Den_d^{-1} \frac{G}{N^c_d}\sum_{t=2}^{F_g-1}N_{g,t}\Delta \Dot{X}_{g,t} E\left(\Delta Y_{g,t}|\bm{D},\bm{X}\right)1\{D_{g,1}=d,F_g\geq 3\}-\theta_d\right).
\end{align*}
Therefore, if we can consistently estimate $E((Y_{g,t}-Y_{g,t-\ell}-(X_{g,t} - X_{g,t-\ell})'\theta_{D_{g,1}})|\bm{D},\bm{X})$ for all $t<F_g$ and $t=F_g-1+\ell$, and $E\left(\Delta Y_{g,t}|\bm{D},\bm{X}\right)$ for all $t<F_g$, then we can consistently estimate $E(U_{g,\ell}|\bm{D},\bm{X})$.
For $t<F_g$,
\begin{align}\label{eq:notgspecific1_X}
&E((Y_{g,t}-Y_{g,t-\ell}-(X_{g,t} - X_{g,t-\ell})'\theta_{D_{g,1}})|\bm{D},\bm{X})\nonumber\\
=&E((Y_{g,t}(\bm{D}_{g,1,t})-Y_{g,t-\ell}(\bm{D}_{g,1,t-\ell}))-(X_{g,t} - X_{g,t-\ell})'\theta_{D_{g,1}})|\bm{D},\bm{X})\nonumber\\
=&\phi(t,\ell,D_{g,1}),
\end{align}
for some function $\phi$. The second equality follows from Assumption \ref{hyp:conds_withX}. Similarly, for $t=F_g-1+\ell$, if one assumes \eqref{eq:constant_effects} and \eqref{eq:onetreatmentchange},
\begin{align}\label{eq:notgspecific2_X}
&E(Y_{g,F_g-1+\ell}-Y_{g,F_g-1}-(X_{g,F_g-1+\ell} - X_{g,F_g-1})'\theta_{D_{g,1}}|\bm{D},\bm{X})\nonumber\\
=&\psi(F_g-1+\ell,\ell,D_{g,1},D_{g,F_g}).
\end{align}
Under Assumption \ref{hyp:conds_withX}, for all $t<F_g$
\begin{align}\label{eq:notgspecific3_X}
E\left(\Delta Y_{g,t}|\bm{D},\bm{X}\right)=\gamma_{t,D_{g,1}}+\Delta X_{g,t}'\theta_{D_{g,1}}. 
\end{align}
In view of \eqref{eq:notgspecific1_X}, \eqref{eq:notgspecific2_X}, and \eqref{eq:notgspecific3_X}, and following similar ideas as those we used for estimators without covariates, let
\[
\widehat{E}_{X,g,t,\ell}^+=
\begin{cases}
\frac{\sum_{g'\in \mathcal{B}^{N,+}_{g,t, \ell}} N_{g',t}[(Y_{g',t}-Y_{g',t-\ell})-(X_{g',t} - X_{g',t-\ell})'\widehat{\theta}_{D_{g,1}}]}{\sum_{g'\in\mathcal{B}^N_{g,t, \ell}}N_{g',t}} & \text{ if } t < F_g \\ 
\noalign{\vskip9pt}
\frac{\sum_{g'\in \mathcal{B}^{S,+}_{g,t, \ell}} N_{g',t}[(Y_{g',t}-Y_{g',t-\ell})-(X_{g',t} - X_{g',t-\ell})'\widehat{\theta}_{D_{g,1}}]}{\sum_{g'\in\mathcal{B}^S_{g,t, \ell}}N_{g',t}} & \text{ if } t = F_g - 1 + \ell, S_g = 1 \\
\noalign{\vskip9pt}
\cdot & \text{ otherwise }
\end{cases}
\]
\[
\widehat{E}_{X,g,t,\ell}^-=
\begin{cases}
\frac{\sum_{g'\in \mathcal{B}^{N,-}_{g,t, \ell}} N_{g',t}[(Y_{g',t}-Y_{g',t-\ell})-(X_{g',t} - X_{g',t-\ell})'\widehat{\theta}_{D_{g,1}}]}{\sum_{g'\in\mathcal{B}^N_{g,t, \ell}}N_{g',t}} & \text{ if } t < F_g \\ 
\noalign{\vskip9pt}
\frac{\sum_{g'\in \mathcal{B}^{S,-}_{g,t, \ell}} N_{g',t}[(Y_{g',t}-Y_{g',t-\ell})-(X_{g',t} - X_{g',t-\ell})'\widehat{\theta}_{D_{g,1}}]}{\sum_{g'\in\mathcal{B}^S_{g,t, \ell}}N_{g',t}} & \text{ if } t = F_g - 1 + \ell, S_g = 0 \\
\noalign{\vskip9pt}
\cdot & \text{ otherwise }
\end{cases}
\]
and for all $t<F_g$ let
$$\widehat{E}\left(\Delta Y_{g,t}|\bm{D},\bm{X}\right)=\left(\widehat{\gamma}_{t,D_{g,1}}+\Delta X_{g,t}'\widehat{\theta}_{D_{g,1}}\right)\ind{\#\{g':D_{g',1}=D_{g,1},F_{g'}>t\}\geq 2},$$ 
where $\widehat{\gamma}_{t,D_{g,1}}$ is the period-$t$ fixed effect in the the OLS regression of $\Delta Y_{g,t}$ on $\Delta X_{g,t}$ and time fixed effects, in the subsample  $ \mathcal{C}^{D_{g,1}}$ (\textit{Code tag: see the ``Running residualization regression'' section in the code.}). Finally, let
\scriptsize
\begin{align*}
&U^{+,var,X}_{g,\ell} =\\
&1\{\ell\leq T_g-1\}\frac{G}{N^1_{\ell}} \sum_{t=\ell+1}^{T_g}\left[\ind{F_g-1+\ell=t,S_g=1}- \frac{N^1_{t,\ell,g}}{N^g_t}\ind{F_g>t}\right] N_{g,t}\text{DOF}^+_{g,t, \ell}\left(Y_{g,t}-Y_{g,t-\ell}-(X_{g,t} - X_{g,t-\ell})'\widehat{\theta}_{D_{g,1}}-\widehat{E}^+_{X,g,t,\ell}\right)\\
&-\sum_{d\in \mathcal{D}^{\text{r}}_1}M^+_{d,\ell}\left[Den_d^{-1} \frac{G}{N^c_d}\sum_{t=2}^{F_g-1}N_{g,t}\Delta \Dot{X}_{g,t}\left(1+\ind{\#\{g':D_{g',1}=d,F_{g'}>t\}\geq 2}\left(\sqrt{\frac{\#\{g':D_{g',1}=d,F_{g'}>t\}}{\#\{g':D_{g',1}=d,F_{g'}>t\}-1}}-1\right)\right)\right.\\
&\left.\left(\Delta Y_{g,t}-\widehat{E}\left(\Delta Y_{g,t}|\bm{D},\bm{X}\right)\right)1\{D_{g,1}=d,F_g\geq 3\}-\widehat{\theta}_d\right]\\
& \\
&U^{-,var,X}_{g,\ell} =\\
&1\{\ell\leq T_g-1\}\frac{G}{N^0_{\ell}} \sum_{t=\ell+1}^{T_g} \left[\frac{N^0_{t,\ell,g}}{N^g_t}\ind{F_g>t}-\ind{F_g-1+\ell=t,S_g=0}\right]N_{g,t}\text{DOF}^-_{g,t, \ell}\left(Y_{g,t}-Y_{g,t-\ell}-(X_{g,t} - X_{g,t-\ell})'\widehat{\theta}_{D_{g,1}}-\widehat{E}^-_{X,g,t, \ell}\right)\\
+&\sum_{d\in \mathcal{D}^{\text{r}}_1}M^-_{d,\ell}\left(Den_d^{-1} \frac{G}{N^c_d}\sum_{t=2}^{F_g-1}N_{g,t}\Delta \Dot{X}_{g,t}\left(1+\ind{\#\{g':D_{g',1}=d,F_{g'}>t\}\geq 2}\left(\sqrt{\frac{\#\{g':D_{g',1}=d,F_{g'}>t\}}{\#\{g':D_{g',1}=d,F_{g'}>t\}-1}}-1\right)\right)\right.\\
&\left.\left(\Delta Y_{g,t}-\widehat{E}\left(\Delta Y_{g,t}|\bm{D},\bm{X}\right)\right)1\{D_{g,1}=d,F_g\geq 3\}-\widehat{\theta}_d\right)\\
& \\
&U^{var,X}_{g,\ell}=  \frac{N^1_\ell}{N^1_\ell+N^0_\ell} U^{+,var,X}_{g,\ell}+  \frac{N^0_\ell}{N^1_\ell+N^0_\ell} U^{-,var,X}_{g,\ell}.
\end{align*}
\normalsize
Our estimator of the variance of $\DID^X_\ell$ is
\begin{align}
\widehat{\sigma}^2_{X,\ell}=&\frac{1}{G^2}\sum_{g=1}^G \left(U^{var,X}_{g,\ell}\right)^2.
\end{align}

\subsection{Allowing for different, potentially non-linear, trends across sets of groups: \st{trends\_nonparam(varlist)}}  
\label{sec:trendsnonparam}

\paragraph{Identifying assumption.}
In some cases, controlling for covariates may be insufficient to account for differences in trends between groups. Then, a common remedy in two-way fixed effect regressions consists in including interactions between time FE and FE for sets of groups. For instance, if groups are US counties, one can allow for state-specific trends. A similar idea can be pursued with the estimators computed by the command. Let $s(g)\in \{1,...,S\}$ denote the set of groups $g$ belongs to. In this set-up, we modify the parallel trends assumption underlying the estimators as follows.
\begin{hyp}\label{hyp:CT_supersets}
	(Parallel trends within sets of groups)
$\forall (g, g’)$, if $s(g)=s(g')$, $D_{g,1} = D_{g’,1} \in \mathcal{D}_1^{\text{r}}$, then $\forall t\geq 2$,
$$E[Y_{g,t}(\bm{D}_{g,1,t}) - Y_{g,t-1}(\bm{D}_{g,1,t-1}) | \bm{D}] = E[Y_{g’,t}(\bm{D}_{g’,1,t}) - Y_{g’,t-1}(\bm{D}_{g’,1,t-1}) | \bm{D}].$$
\end{hyp}
Assumption \ref{hyp:CT_supersets} is a weakening of Assumption \ref{hyp:strong_exogeneity}, as it only requires that the status-quo potential outcome of groups in the same set of groups follow the same evolution over time.

\paragraph{Estimation.}
The estimators computed by the command when the \st{trends\_nonparam} option is specified are very similar to the baseline estimators, except that switchers' outcome evolution is compared to the outcome evolution of groups whose treatment has not changed yet, and that belong to the same supergroup $s(g).$ For instance, the outcome evolution of counties whose treatment has changed is compared to the outcome evolution of counties whose treatment has not changed and that belong to the same state. To define those estimators, we just need to redefine some of the objects defined in Section \ref{sec:baseline}. For every $g$, let $$T_g=\max_{g':D_{g',1}=D_{g,1}, \mathcal{S}_{g'}=\mathcal{S}_g}F_{g'}-1$$ denote the last period when there is still a group from the same supergroup as $g$, with the same treatment as $g$'s in period one, and whose treatment has not changed since the start of the panel. For all $(g,t)$, let $$N^g_t=\sum_{g':D_{g',1}=D_{g,1},\mathcal{S}_{g'}=\mathcal{S}_g,F_{g'} >t}N_{g',t}$$ denote the number of observations at period $t$ in groups $g'$ from the same supergroup as $g$, with the same period-one treatment as $g$, and that kept the same treatment from period $1$ to $t$. The definitions of $S_g$, $L_u$, $N_\ell^1$, $w_{+,\ell}$, $\delta^D_{+,\ell}$, $L_a$, $N_\ell^0$, $w_{-,\ell}$, and $\delta^D_{-,\ell}$  remain the same as before. To estimate $\delta_{g,\ell}$, we use
\begin{equation*}
\DID_{g,\ell} =Y_{g,F_g-1+\ell} - Y_{g,F_g-1} -
\sum_{g':D_{g',1}=D_{g,1},\mathcal{S}_{g'}=\mathcal{S}_g,F_{g'} >F_g-1+\ell} \frac{N_{g',F_g-1+\ell}}{N^{g}_{F_g-1+\ell}}(Y_{g',F_g-1+\ell} - Y_{g',F_g-1}).
\end{equation*}
Then, the definitions of $\DID_{+,\ell}$, $\DID_{-,\ell}$, and $\DID_{\ell}$ remain the same as before.
For every $\ell \in \{1,...,L_u\}$, $t\in \{\ell+1,...,T_g\}$, and $g$, let
$$N^1_{t,\ell,g}=\sum_{g':S_{g'}=1, F_{g'}-1+\ell=t,D_{g',1}=D_{g,1},\mathcal{S}_{g'}=\mathcal{S}_g, N^{g'}_t>0} N_{g',t}.$$ 
For every $\ell \in \{1,...,L_a\}$, $t\in \{\ell+1,...,T_g\}$, and $g$, let
$$N^0_{t,\ell,g}=\sum_{g':S_{g'}=0, F_{g'}-1+\ell=t,D_{g',1}=D_{g,1},\mathcal{S}_{g'}=\mathcal{S}_g, N^{g'}_t>0} N_{g',t}.$$ The definitions of $U^+_{g,\ell}$, $U^-_{g,\ell}$, and $U_{g,\ell}$ remain the same as before.
To estimate the variance of $\DID_{\ell}$, we assume that the number of distinct values of $(D_{g,1},F_g,D_{g,F_g},\mathcal{S}_g)_{g: D_{g,1}\in \mathcal{D}^{\text{r}}_1}$ is finite.
Let $f(g) = (D_{g,1},F_g,D_{g,F_g},\mathcal{S}_g)$ for all $g \in \mathcal{G}$. Then, the definitions of $\mathcal{C}_g$,  $\mathcal{N}_{d,t}$, $\mathcal{S}_{d,t,\ell}$, $\mathcal{U}_{d,t,\ell}$, $\widehat{E}_{g,t, \ell}$, $U^{+,var}_{g,\ell}$, $U^{-,var}_{g,\ell}$, $U^{var}_{g,\ell}$, and $\widehat{\sigma}^2_\ell$ remain the same as before.

\paragraph{Combining the \st{trends\_nonparam(varlist)} and  \st{controls(varlist)} options.}
When those two options are combined, the identifying assumption underlying estimators becomes:
there are vectors $(\theta_d)_{d\in\mathcal{D}_1^{\text{r}}}$ of same dimension as $X_{g,t}$ such that 
$\forall (g, g’)$, if $s(g)=s(g')$, $D_{g,1} = D_{g’,1} \in \mathcal{D}_1^{\text{r}}$, then $\forall t\geq 2$,
\begin{align}
& E[Y_{g,t}(\bm{D}_{g,1,t}) - Y_{g,t-1}(\bm{D}_{g,1,t-1}) - (X_{g,t} - X_{g,t-1})'\theta_{D_{g,1}} | \bm{D},\bm{X}] \notag\\
= & E[Y_{g’,t}(\bm{D}_{g’,1,t}) - Y_{g’,t-1}(\bm{D}_{g’,1,t-1}) - (X_{g',t} - X_{g',t-1})'\theta_{D_{g',1}} | \bm{D},\bm{X}].\label{eq:CT_supergroups_controls} 
\end{align}
Thus, the residualization of the outcome with respect to the controls still takes places within values of the baseline treatment, and not within  values of the baseline treatment $\times$ sets of groups.   

\subsection{Allowing for group-specific linear trends: \st{trends\_lin}}  
\label{sec:trends_lin}

When this option is specified, the command starts by computing the event-study estimators described above, with an outcome variable equal to the outcome's first-difference rather than to the outcome itself. Those estimators rely on a parallel trends assumption on the outcome's first-difference, thus allowing for group-specific linear trends. However, they are unbiased for event-study effects the outcome's first-difference. To recover event-study effect $\ell$ on the outcome, event-study effects on the outcome's first-difference are summed from 1 to $\ell$. One needs to ensure that all event-study effects on the outcome's first-difference apply to the same switchers, as otherwise the sum of those event-study effects would not apply to a well-defined population of switchers. See Section 1.3 of the paper's Web Appendix for further details. 

\subsection{Estimation with a continuous period-one treatment: \st{continuous(\#)}}\label{sec:continuous}

\paragraph{Identifying assumption.}
The command can be used when groups' period-one treatment is continuous, meaning that all groups have a different period-one treatment value. With a discrete period-one treatment, the command compares the outcome evolution of switchers and non-switchers with the same period-one treatment. 
But with a truly continuous period-one treatment, there will be no two groups with the same period-one 
treatment. Then, the command assumes that group's status-quo outcome evolution is a polynomial in their period-one treatment. 
Specifically,
\begin{hyp}
	(Functional parallel trends for the status-quo outcome, conditional on the period-one treatment) For all $t\ge 2$, $E[Y_{g,t}(\bm{D}_{g,1,t}) - Y_{g,t-1}(\bm{D}_{g,1,t-1}) | \bm{D}]=\sum_{k=0}^K \gamma_{k,t} D_{g,1}^k$.
	\label{hyp:strong_exogeneity3}
\end{hyp}
The assumed polynomial order $K$ is the option's argument.

\paragraph{Estimators.}
To compute the non-normalized event-study estimators with the \st{continuous(\#)} option, the command starts by redefining the treatment as $\tilde{D}_{g,t}=1\{S_g=1,t\geq F_g\}-1\{S_g=0,t\geq F_g\}$, a variable equal to zero before groups' treatment switches, equal to $1$ for switchers-in after they switch, and to $-1$ for switchers-out after they switch. Then, the estimation is run with this modified treatment, adding as control variables
$(D_{g,1}^k1\{t\geq t'\})_{k\in\{0,...,K\},t'\in \{1,...,T\}}$, a full set of interactions between the polynomial terms in the baseline treatment and indicators for period $t$ being after $t'$. As $\tilde{D}_{g,1}=0$ for all $g$, the command runs only one residualization regression of $Y_{g,t}-Y_{g,t-1}$ on $(D_{g,1}^k1\{t\geq t'\})_{k\in\{0,...,K\},t'\in \{1,...,T\}}$ (and potentially other controls specified by the user), among all $(g,t)$ cells whose treatment has not changed yet at $t$. To compute the average total effect or normalized event-study estimators, the resulting non-normalized event-study estimators are used in the numerator, and the denominator is computed using the formulas given above, but with the original treatment $D_{g,t}$ instead of  $\tilde{D}_{g,t}$. The estimators produced with the 
\st{continuous(\#)} option are unbiased under Assumption \ref{hyp:strong_exogeneity3}, if the requested polynomial order is greater than or equal to $K$.

\paragraph{Variance estimators.}
Unlike the other variance estimators computed by the command, those computed when the \st{continuous} option is specified are not backed by a proven asymptotic normality result, or at least by a linearization of the estimators. Preliminary simulation evidence indicates that when the option is used with a correctly-specified polynomial order, those variance estimators are conservative. On the other hand, when the specified polynomial order is strictly larger than that in Assumption \ref{hyp:strong_exogeneity3}, those variance estimators can become liberal. Thus, when this option is specified, we recommend using the bootstrap for inference, by manually bootstrapping the command. At least, one should perform a robustness check where one compares the analytic variance computed by the command to a bootstrapped variance.  Soon, the command will have an option to automatically perform this bootstrap when the \st{continuous} option is specified.     

\subsection{Clustering standard errors: \st{cluster(varlist)}}\label{sec:clustering}

The command can compute variance estimators clustered at a coarser level than groups. Then, the variance estimators are not very different from those given in the definition of $\widehat{\sigma}^2_\ell$ above, except that the variables $U^{var}_{g,\ell}$ are summed across groups in a given cluster, and the resulting variable is squared and summed across clusters  (\textit{Code tag: see the ``sum U\_Gg\_var\_l within a cluster'' section in the code}). Another difference is that with clustering, cohorts' cardinality (e.g.: $\#\mathcal{U}_{d,t,\ell}$) are now defined as the number of different clusters that the groups in the cohort belong to. Consequently, while $\#\mathcal{U}_{d,t,\ell}>1$ without clustering, with clustering we could have $\#\mathcal{U}_{d,t,\ell}=1:$ all switchers and not yet switchers with the same baseline treatment may belong to the same cluster. When $\#\mathcal{U}_{d,t,\ell}=1$, we let $\widehat{E}_{g,t}=0$ and we let $\text{DOF}_{g,t}=1$.  

\subsection{Estimating heterogeneous effects: \st{by(\textit{varname})} and \st{predict\_het(\textit{varlist,numlist})}}  
\label{sec:hetX}

When the option \st{by(\textit{varname})} is specified, the command estimates all effects separately by the levels of \textit{varname}, a group-level and time-invariant variable.  If \textit{varname} is a binary variable for example, then the estimation is carried out once for groups with \textit{varname}=0 and once for groups with \textit{varname}=1. Then, the command reports on a graph event-study plots for all values of \textit{varname}, thus allowing to assess effect heterogeneity by \textit{varname}.

\medskip
When the option \st{predict\_het(varlist,numlist)} is specified, the command outputs tables showing whether the group-level and time-invariant variables in \textit{varlist} predict groups' estimated event-study effects. By default, with this option the
    command produces one table per event-study effect estimated, each displaying the coefficients from regressions of the group-level estimate of the event-study effect on the variables in \textit{varlist}. The p-value of a test that all coefficients are
    equal to zero is shown below each table. If one is only interested in predicting a subset of the event-study effects estimated, one can specify that subset inside \textit{numlist}. This option cannot be specified together with the \st{normalized} or \st{controls} options. It builds on ideas proposed by \cite{shahn2023subgroup} and \cite{dCDHtextbook}. See Section 1.5 of the Web Appendix of de Chaisemartin and D'Haultfoeuille (2024) for further details. 

\subsection{Ensuring that all event-study effects are estimated for the same switchers: \st{same\_switchers}}  
\label{sec:same_switchers}

The sample of switchers to which $\DID_\ell$ or $\DID^n_\ell$ apply changes with $\ell$, so compositional changes can account for the variation in those effects across $\ell$. This issue is present in all event-study estimation methods, and is not specific to  
\st{did\_multiplegt\_dyn}. The \st{same\_switchers} option can be used to avoid such compositional changes: when it is specified, effects are estimated only for switchers for which all the requested effects can be estimated. If the \st{same\_switchers\_pl} option is also specified, then 
placebos are estimated only for switchers for which all the requested effects and placebos can be estimated. \st{same\_switchers\_pl} can only be specified if \st{same\_switchers} is also specified.  (\textit{Code tag: see the ``If the same\_switchers option is specified'' section in the code.})

\subsection{Estimating event-study effects separately for switchers in/out: \st{switchers(\textit{in/out})}}  
\label{sec:in_out}

One may be interested in estimating separately the treatment effect of switchers-in, whose treatment after they switch is larger than their period-one treatment, and of switchers-out, whose treatment after they switch is lower than their period-one treatment.  In that case, one should run the command first with the \st{switchers(in)} option, and then with the \st{switchers(out)} option.

\section{Simulations}
\label{sec:simulations}

\subsection{Simulations based on the data of \cite{vella1998whose}}

\paragraph{Data.}
Our first simulations use the data from \cite{vella1998whose}, a balanced panel of 545 young American workers, which contains their wages and their union status each year from 1980 to 1987.

\paragraph{Design.}
We begin with a simple binary and staggered design. Let $F_g$ denote the first date when the union status of worker $g$ is different from its period-one union status. Then, we let $D_{g,t}=1\{t\ge F_g\}$.

\paragraph{Outcome model.}
In each simulation, worker $g$ is assigned a vector of outcome shocks $$(\eps_{g,2},…,\eps_{g,T}),$$ drawn randomly from workers' actual first-differenced wages $$(Y_{g',2}-Y_{g',1},…,Y_{g',T}-Y_{g',T-1})_{g'\in \{1,...,545\}}.$$ Then, the simulated status-quo outcome of worker $g$ at $t$ is
\begin{equation}\label{eq:model_sqo}
Y_{g,t}(\bm{D}_{g,1,t})=Y_{g,1}+\sum_{k=2}^t \eps_{g,k}+t^2.    
\end{equation}
Finally, we let the observed outcome be 
\begin{equation}\label{eq:obsY}
Y_{g,t}=Y_{g,t}(\bm{D}_{g,1,t}),    
\end{equation}
meaning that the current and lagged treatments do not have any effect on the outcome. 

\paragraph{Results.}
We generate 2000 data sets as described above, and on each simulated data we run \\ \st{did\_multiplegt\_dyn Y G T D, effects(3) placebo(3),} \\
and compute the coverage rate of the confidence intervals (CIs) of the three estimated effects and placebos. We also compute the rejection rate of the F-test that all placebos are jointly equal to zero. The first line of Table \ref{tbl:sim_main} below shows the results. CIs' coverage rates are very close to their nominal 0.95 level, for all effects and for all placebos. The rejection rate of the F-test that all placebos are equal to zero is also very close to 0.05.

\paragraph{Simulations with lower values of $G$.}
In the second to fifth lines of  Table \ref{tbl:sim_main}, we replicate those simulations, in random samples of 100, 50, 30, and 20 workers, each time with 50\% of switchers and non-switchers. CIs' coverage rates remain very good with as few as 50 workers, and even with 20 workers their coverage rate is never below 0.919. The F-test starts over-rejecting with 50 workers, and the over-rejection is substantial with 20 workers. 


\paragraph{Simulations with a treatment effect.}
The sixth line of Table \ref{tbl:sim_main} shows simulations where we replace \eqref{eq:model_sqo} by
$Y_{g,t}=Y_{g,t}(\bm{D}_{g,1,t})+2\times (D_{g,t}-D_{g,1}):$ one unit of the current treatment increases the outcome by two units, but lagged treatments do not affect the outcome.
CIs' coverage rates remain very close to their nominal level, for all effects and for all placebos. In this DGP, the non-normalized effects $\delta_\ell$ are all equal to 2. We use the command's \st{effects\_equal} option to test that effects are equal. As expected, the test's rejection rate is close to $0.05$.
 
\paragraph{Simulations where we estimate normalized effects.}
The seventh line of Table \ref{tbl:sim_main} shows simulations with the exact same DGP as in the first line, but where we run \\ 
\st{did\_multiplegt\_dyn Y G T D, effects(3) placebo(3) normalized,} \\
to estimate normalized event-study effects. With $Y_{g,t}=Y_{g,t}(\bm{D}_{g,1,t})+2\times (D_{g,t}-D_{g,1})$, and in designs where groups' treatment can only change once, normalized event-study effect $\ell$ is equal to $2/(\ell+1)$. CIs' coverage rates are identical to those in the previous line, because normalized estimators and standard errors are equal to non-normalized ones, divided by the same constant.

\paragraph{Simulations with control variables.}
The eighth line of Table \ref{tbl:sim_main} shows simulations where instead of \eqref{eq:model_sqo}, we assume that 
\begin{equation*}
Y_{g,t}(\bm{D}_{g,1,t})=Y_{g,1}+\sum_{k=2}^t \eps_{g,k}+t^2+ \frac{\sigma}{\mu} \text{hours}_{g,t} + 2\sigma\times \text{married}_{g,t},   \end{equation*}
where $\sigma$ is the sample standard deviation of wages  and $\mu$ the average of hours worked.
In this DGP, Assumption \ref{hyp:strong_exogeneity} fails, but Assumption \ref{hyp:conds_withX} holds with  control variables \st{hours} and \st{married}. Then, we run \\
\st{did\_multiplegt\_dyn Y G T D, effects(3) placebo(3) normalized controls(hours married).} \\
CIs' coverage rates remain very close to their nominal level, for all effects and for all placebos. The rejection rate of the F-test remains close to 0.05.

\paragraph{Simulations with education-specific non-linear trends.}
Let $\text{educ}_g$ be equal to $1$ if worker $g$ has less than $12$ years of education; $2$ if they have exactly $12$ years; and $3$ if they have more than $12$ years. The ninth line of Table \ref{tbl:sim_main}
shows simulations where instead of \eqref{eq:model_sqo}, we assume that 
\begin{equation*}
Y_{g,t}(\bm{D}_{g,1,t})=Y_{g,1}+\sum_{k=2}^t \eps_{g,k}+t^2\times \text{educ}_g,    
\end{equation*}
so that Assumption \ref{hyp:strong_exogeneity} fails, but Assumption \ref{hyp:CT_supersets} holds with $s(g)=\text{educ}_g$. Then, we run \\ \st{did\_multiplegt\_dyn Y G T D, effects(3) placebo(3) normalized trends\_nonparam(educ).} \\
CIs' coverage rates remain very close to their nominal level, for all effects and for all placebos. The rejection rate of the F-test remains close to 0.05.

\paragraph{Simulations with education-specific non-linear trends and control variables.}
Line 10 of Table \ref{tbl:sim_main}
shows simulations where instead of \eqref{eq:model_sqo}, we assume that 
\begin{equation*}
Y_{g,t}(\bm{D}_{g,1,t})=Y_{g,1}+\sum_{k=2}^t \eps_{g,k}+t^2\times \text{educ}_g+\frac{\sigma}{\mu} \text{hours}_{g,t} + 2\sigma\times \text{married}_{g,t},    
\end{equation*}
so that Assumption \ref{hyp:strong_exogeneity} fails, but \eqref{eq:CT_supergroups_controls} holds with $s(g)=\text{educ}_g$ and  control variables \st{hours} and \st{married}.
Then, we run \\ \st{did\_multiplegt\_dyn Y G T D, effects(3) placebo(3) normalized trends\_nonparam(educ)  controls(hours married).} \\
CIs' coverage rates remain very close to their nominal level, for all effects and for all placebos. The rejection rate of the F-test remains close to 0.05.

\paragraph{Simulations with worker-specific linear trends.}
Line 11 of Table \ref{tbl:sim_main} 
shows simulations where instead of \eqref{eq:model_sqo}, we assume that 
\begin{equation*}
Y_{g,t}(\bm{D}_{g,1,t})=Y_{g,1}+\sum_{k=2}^t \eps_{g,k}+t^2+t\times \text{trend}_g,    
\end{equation*}
where $\text{trend}_g = \sum_{k=1}^5Q^k(Y_{g,1})$, and $Q^k(Y_{g,1})$ is a dummy equal to 1 if the baseline wage of worker g is higher than the $k$-th quintile of the distribution of wages at period one. Then, we run \\ \st{did\_multiplegt\_dyn Y G T D, effects(3) placebo(3) normalized trends\_lin} \\
Here, the coverage for the average total effect and third placebo are missing. This is because \st{did\_multiplegt\_dyn} does not estimate the average total effect when the \st{trends\_lin} option is specified. Moreover, as \st{trends\_lin} relies on first-differencing the outcome, when that option is specified there are not enough pre-treatment periods to estimate the third placebo. CIs' coverage rates remain very close to their nominal level, for all effects and for all placebos. The rejection rate of the F-test remains close to 0.05.

\paragraph{Simulations with clustering.} Line $12$ of Table \ref{tbl:sim_main} shows simulations where we perform the following procedure. First, we keep $500$ workers, those with the lowest wage at period one. Then we create 50 clusters of 10 workers as follows: the first cluster groups the 10 workers with lowest period one wages, the second cluster groups the next 10 workers with lowest wages, and so forth. Let the clusters be indexed by $c$, and let $i$ represent the numbering of workers within clusters ($1$ for the one with the lowest wage, ... $10$ for the one with the highest wage). Next, let 
\begin{equation*}
Y_{i,c,t}(\bm{D}_{i,c,1})=Y_{i,c,1}+\sum_{k=2}^t \eps_{i,c,k}+t^2 + \eta_{c,t}.  
\end{equation*}
$\eta_{c,t}$ is a cluster specific AR(1) process, constructed as:
\begin{equation*}
    \eta_{c,t}  = \frac{1}{\sqrt{2}}\eta_{c,t-1} + \frac{1}{\sqrt{2}}\nu_{c,t}
\end{equation*}
where $\nu_{c,t}\sim\mathcal{N}(0,V^{2})$ iid, and $V=\hat{\mathbb{V}}(Y_{g,2}-Y_{g,1})$. Finally, we draw the residuals in a clustered manner. In each simulation, each worker $i$ from cluster $c$ is assigned a vector of outcome shocks $(\eps_{i,c,2},…,\eps_{i,c,T}),$ 
where 
$(\eps_{i,c,2},…,\eps_{i,c,T})_{i\in \{1,...,10\}}$ is
drawn randomly from clusters' actual wage's first differences $$((Y_{i,c',2}-Y_{i,c',1},…,Y_{i,c',T}-Y_{i,c',T-1})_{i\in \{1,...,10\}})_{c'\in \{1,...,50\}}.$$ 
The cluster-specific shocks $\eta_{c,t}$ and the clustered draw of the worker-specific shocks $(\eps_{i,c,2},…,\eps_{i,c,T})$ create substantial correlation of $Y_{i,c,t}(\bm{D}_{i,c,1})$ within clusters. For instance, the intra-cluster correlation coefficient of $$1/T\sum_{t=1}^TY_{i,c,t}(\bm{D}_{i,c,1})$$ is equal to $0.827$. Therefore, Assumption \ref{hyp:indep_groups} fails. On the other hand, $(Y_{i,c,1}(\bm{D}_{i,c,1}),...,Y_{i,c,T}(\bm{D}_{i,c,1}))_{i\in \{1,...,10\}}$ is independent across clusters. Therefore, we run \\ \st{did\_multiplegt\_dyn Y G T D, effects(3) normalized placebo(3) cluster(cluster)} \\
where \st{cluster} is the cluster of worker $g$. CIs' coverage rates remain very close to their nominal level, for all effects and for all placebos. The F-test slightly over-rejects.

\paragraph{Simulations with a binary non-staggered treatment.}
In Line 13 of Table \ref{tbl:sim_main}, instead of $D_{g,t}$, we redefine the treatment as worker $g$'s union status at $t$. This yields a binary and non-staggered treatment: workers can join but also leave the union, and can do so multiple times. Status quo potential outcomes are generated following \eqref{eq:model_sqo}, and we run \\ 
\st{did\_multiplegt\_dyn Y G T D, effects(3) placebo(3) normalized more\_granular\_demeaning.} \\
The \st{more\_granular\_demeaning} option ensures our variance estimators are not conservative, given that workers' treatment can change more than once. CIs' coverage rates remain very close to their nominal level, for all effects and for all placebos. The rejection rate of the F-test remains close to 0.05.

\subsection{Simulations based on the data of \cite{wolfers2006did}}

\paragraph{Dataset}
To ensure that our results are not specific to the \cite{vella1998whose} dataset, we also conduct simulations using the dataset of \cite{wolfers2006did}, a panel data set of states in the U.S. ranging from $1956$ to $1986$, where the outcome variable $Y_{g,t}$ is the divorce rates per $1000$ people. The treatment variable $D_{g,t}$ is whether state $g$ has passed a unilateral divorce law in year $t$, a binary and absorbing treatment. The panel is unbalanced because there are states whose divorce rate is not observed in every year, but we focus on the restricted sample of 42 states whose divorce rate is observed in every year. We further drop two states that are already treated in 1956: as the treatment is binary and absorbing, those states will not contribute to the estimators computed by the \st{did\_multiplegt\_dyn} command. Thus, we are left with 40 states, 30 of which adopt the treatment over the study period, while 10 remain untreated.

\paragraph{Outcome model.}
Analogously to the previous simulations, each state is assigned to a vector of outcome shocks drawn randomly from the states' actual divorce rate's first difference. Moreover, the simulated potential outcomes are given by
\begin{equation}\label{eq:model_wolfers}
Y_{g,t}(\bm{D}_{g,1956,t})=Y_{g,1956}+\sum_{k=1956}^t \eps_{g,k}+t^2.   
\end{equation}
The observed outcome follows
\begin{equation}
    Y_{g,t} = Y_{g,t}(\bm{D}_{g,1956,t}),
\end{equation}
meaning that the treatment has no effect. 

\paragraph{Simulations with the original treatment variable.} In the first line of Panel B, we generate 2000 data sets as described above, and on each simulated data we run \\ \st{did\_multiplegt\_dyn Y G T D, effects(3) placebo(3),} \\
and compute the coverage rate of the confidence intervals (CIs) of the three estimated effects and placebos. We also compute the rejection rate of the F-test that all placebos are jointly equal to zero. Line 1 of Panel $B$ of Table \ref{tbl:sim_main} shows the results. CIs' coverage rates are very close to their nominal level, for all effects and for all placebos. The rejection rate of the F-test is close to 0.05.

\paragraph{Simulations with a recoded  treatment with one state per cohort.}
The original treatment variable is such that 1 state adopts in 1969, 2 states adopt in 1970, 6 states adopt in 1971, 2 states adopt in 1972, 8 states adopt in 1973, 2 states adopt in 1974, 1 state adopts in 1975, 1 state adopts in 1976, 1 state adopts in 1977, and 1 state adopts in 1985. In the second line of Panel B we recode the treatment variable so that we have exactly one state becoming treated in each year$\geq 1957$. This yields a panel of 40 states where 30 of them are treated, each on a different date. CIs' coverage rates are now slightly larger than their nominal level, around 0.97, thus suggesting that the command's CIs are around 2.17/1.96=10.7\% too large. Accordingly, the rejection rate of the F-test is much lower than 0.05, around 0.006.

\paragraph{Simulations with a recoded  treatment with one state per cohort, and a treatment effect.} The DGP in the third line is the same as in the second, except that we add a constant effect of the current treatment on the outcome equal to 0.25, calibrated to the estimated short-run effect of the treatment in this application \citep{de2022survey}. Then, we let
\begin{equation}
    Y_{g,t} = Y_{g,t}(\bm{D}_{g,1956,t})+0.25D_{g,t}.
\end{equation}
As expected, with one state per cohort, our CIs for the dynamic effects become more conservative with a treatment effect: now their coverage rate is around $0.99$, thus suggesting that the command's CIs are around 2.58/1.96=31.6\% too large. 


\subsection{Simulations based on the data of \cite{gentzkow2011}}

Our two previous datasets have a binary treatment and a balanced panel. To investigate the coverage rate of our CIs with a non-binary treatment and an imbalanced panel, in Panel C of Table \ref{tbl:sim_main} we run simulations based on the data of \cite{gentzkow2011}. This is a US county-level panel data set, with 1195 counties, where the data is measured every four years, in each US presidential election. We use 8 elections, from 1900 to 1928. The outcome variable $Y_{g,t}$ is the turnout rate in county $g$ and election-year $t$. The treatment variable $D_{g,t}$ is county $g$'s number of newspapers at $t$, a discrete non-binary variable. The sample has $9239$ observations: either $Y_{g,t}$ or $D_{g,t}$ is missing for $321$ observations, thus resulting in an imbalanced panel. To simplify the design, we cap the treatment at 4, and after a county's number of newspapers has changed once, we assume it keeps the same treatment throughout. In each simulation, we generate vectors of outcome shocks $(\eps_{g,2},…,\eps_{g,T})$ as we did with the previous datasets, and then use \eqref{eq:model_sqo} and \eqref{eq:obsY} to generate the outcome. CIs' coverage rates are very close to their nominal level, for all effects and for all placebos. The rejection rate of the F-test is close to 0.05.

\begin{landscape}
\begin{center}
\begin{table}[h!]
\caption{Simulation Results \st{did\_multiplegt\_dyn}}
\label{tbl:sim_main}
\scriptsize
\resizebox{\linewidth}{!}{
\begin{tabular}{lcccccccc}
\toprule
 & $\DID_{1}$ & $\DID_{2}$ & $\DID_{3}$ & $\widehat{\delta}$ & $\DID^{\text{pl}}_{1}$ & $\DID^{\text{pl}}_{2}$ & $\DID^{\text{pl}}_{3}$ & F-test placebo \\
\midrule

\midrule
\multicolumn{9}{l}{\textbf{Panel A. Simulations based on the data of \cite{vella1998whose}}} \\
\midrule
Baseline & 0.9515 & 0.956  & 0.9515 & 0.9565 & 0.9545 & 0.953  & 0.9485 & 0.0525 \\
$G=100$ & 0.9485 & 0.953  & 0.9465 & 0.9415 & 0.952  & 0.956  & 0.945  & 0.067 \\
$G=50$  & 0.9425 & 0.9325 & 0.943  & 0.9395 & 0.9365 & 0.9385 & 0.932  & 0.0775 \\
$G=30$  & 0.939  & 0.933  & 0.9405 & 0.9285 & 0.9365 & 0.9455 & 0.9295 & 0.106 \\
$G=20$  & 0.919  & 0.9275 & 0.927  & 0.9275 & 0.9185 & 0.9265 & 0.923  & 0.185 \\
Treatment Effect & 0.9515 & 0.956  & 0.9515 & 0.9565 & 0.9545 & 0.953  & 0.9485 & 0.0525 \\
Normalized       & 0.9515 & 0.956    & 0.952      & 0.9565 & 0.9545 & 0.953  & 0.9485 & 0.0525 \\
Controls         & 0.953  & 0.953  & 0.95   & 0.9555 & 0.9555 & 0.955  & 0.9505 & 0.05 \\
Non-linear trends by educ & 0.9525 & 0.9565 & 0.9495 & 0.9565 & 0.9525 & 0.9545 & 0.952  & 0.0535 \\
Non-linear trends by educ $+$ Controls & 0.9545 & 0.961  & 0.9485 & 0.957  & 0.954  & 0.952  & 0.9495 & 0.054 \\
Linear Trends    & 0.9525 & 0.958  & 0.9635 &      & 0.945  & 0.951  &   & 0.057 \\
Clustered SEs    & 0.951  & 0.9505 & 0.951  & 0.9505 & 0.939  & 0.944  & 0.9355 & 0.085 \\
Non-staggered design & 0.9505 & 0.953  & 0.968  & 0.956  & 0.9545 & 0.9505 & 0.9425 & 0.055 \\

\midrule
\multicolumn{9}{l}{\textbf{Panel B. Simulations based on the data of \cite{wolfers2006did}, balanced subsample}} \\
\midrule
Baseline & 0.959  & 0.951  & 0.948  & 0.964  & 0.968  & 0.963  & 0.96   & 0.07 \\
1 switcher / cohort & 0.9715 & 0.9725 & 0.968  & 0.9735 & 0.9725 & 0.966  & 0.968  & 0.0065 \\
1 switcher / cohort, $TE=0.25$ & 0.9965 & 0.99   & 0.984  & 0.9905 & 0.9725 & 0.966  & 0.968  & 0.0065 \\

\midrule
\multicolumn{9}{l}{\textbf{Panel C. Simulations based on the data of \cite{gentzkow2011}}} \\
\midrule
Baseline & 0.95   & 0.9465 & 0.9555 & 0.949  & 0.945  & 0.9505 & 0.9495 & 0.057 \\

\bottomrule
\end{tabular}
}
\end{table}
\end{center}
\end{landscape}

\section{Four Examples Based on Real
Datasets}

In this section, the Stata command \texttt{did\_multiplegt\_dyn} is used on real datasets to  estimate event-study effects in complex designs. The first example has a binary treatment that can turn on an off. The second example has a continuous absorbing treatment. The third example has a discrete multivalued treatment that can increase or decrease multiple times over time. The fourth example has two, binary and absorbing treatments, where the second treatment always happens after the first. To access the datasets used in the following examples, first run: 
\begin{lstlisting}
ssc install did_multiplegt_dyn
net get did_multiplegt_dyn
\end{lstlisting}

\subsection{\citet{Deryugina2017}: On and Off Binary Treatment}

\subsubsection{Research Question}

How do hurricanes affect non-disaster government transfer programs in the United States?

\subsubsection{Data}

\begin{lstlisting}
use Deryugina_2017.dta, clear
\end{lstlisting}
The data come from the publicly available dataset in the replication package of \cite{deryugina_dataset2017}. The data is a panel of US counties at the yearly level. The variable \texttt{hurricane} is the binary treatment. The outcome is \texttt{log\_curr\_trans\_ind\_gov\_pc}, the log of non-disaster government transfer received by county $g$ in year $t$, divided by that county's population.

\subsubsection{On-and-off one-shot design: counties are treated for at most one period}

\begin{lstlisting}
egen total_hurricane = total(hurricane), by(county_fips)
sum total_hurricane 
\end{lstlisting}
We sum the treatment values across years by county, and the total treatment takes the value of 0 or 1 for all counties. Therefore, we are working with a non-absorbing binary treatment, that can turn on and off. Moreover, counties are treated for at most one year, so we are in a so-called ``one-shot'' design \citep{de2022difference}.

\subsubsection{Estimate Treatement Effects with \texttt{did\_multiplegt\_dyn}}

The author uses an event-study two-way fixed effects regression to estimate hurricanes' effects on non-disaster government transfer per capita for 11 years after a hurricane. Instead, we use \texttt{did\_multiplegt\_dyn} to estimate the same effects with an estimator robust to heterogeneous effects across counties and over time. We also compute eleven pre-trends estimators. 

\begin{lstlisting}
* Call did_multiplegt_dyn
did_multiplegt_dyn log_curr_trans_ind_gov_pc county_fips year hurricane, effects(11) placebo(11) cluster(county_fips)
\end{lstlisting}
\begin{figure}[H]
    \centering
    \includegraphics[width=0.7\textwidth]{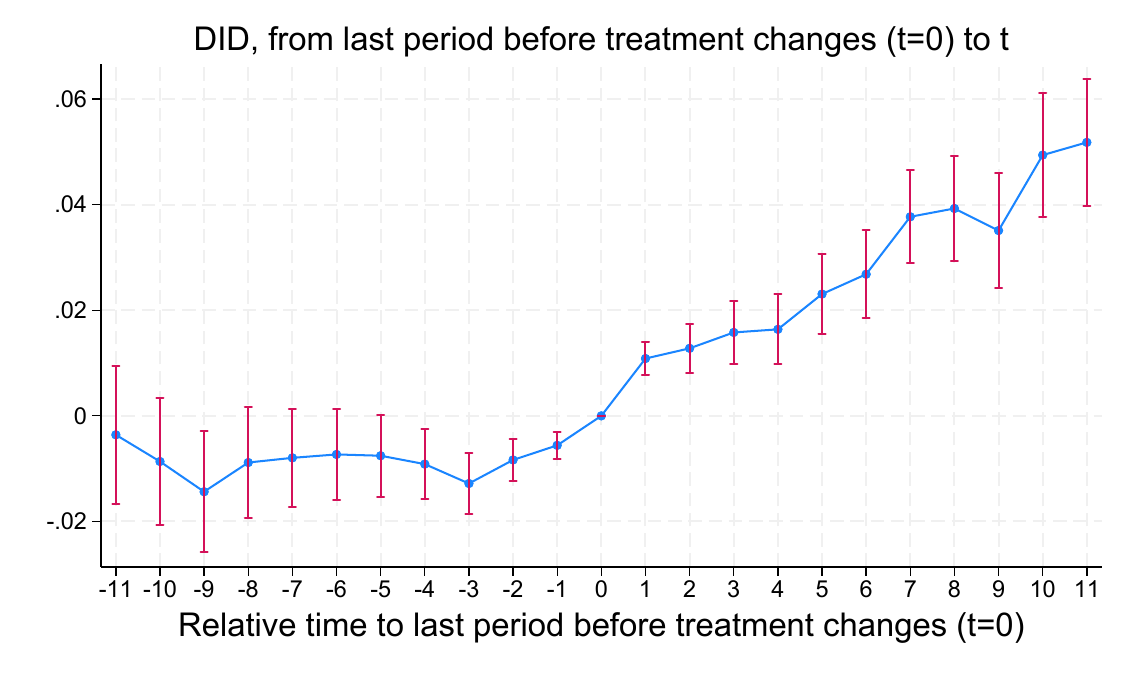}  
    \caption{\texttt{did\_multiplegt\_dyn}: Effects of Hurricane Hits on Government Transfers Per Capita}
    \label{fig:yourlabel}
\end{figure}

The event-study estimators suggest that hurricanes increase non-disaster government transfer per capita by 2\% in the short run, and by 4\% in the long run. In a one-shot treatment design, under the parallel-trends assumption event-study effects significantly different from zero at $\ell \geq 2$ imply that we can reject the no-dynamic or no-carryover effect hypothesis \citep{liu2024practical}. Intuitively, as treated groups revert back to being untreated immediately after receiving the treatment, any remaining treatment effect in subsequent periods must come from effects of lagged treatments on the outcome.

\medskip
The previous discussion relies on no-anticipation and parallel-trends assumptions. However, the pre-trend estimators are statistically significant, and the event-study estimators could just be the continuation of differential trends between treated and untreated counties that seem to start 4 (-3-1) periods before the treated counties get treated.   

\medskip
To address this potential violation of the no-anticipation and parallel-trends assumptions, we follow the author and control for several county-level characteristics. These are all time-invariant variables, like the counties' log population in 1969, that capture demographic and economic conditions at the start of the panel.  Following Section 4.1 of \cite{de2023credible}, the variables inputted to the \texttt{controls()} option of \texttt{did\_multiplegt\_dyn}  have to be defined as $X_g \times t$, to allow for linear trends interacted with the covariates.

\begin{lstlisting}
* Time-invariate covariates: Adding covariates*year
local vars coastal land_area1970 log_pop1969 frac_young1969 frac_old1969 frac_black1969 log_wage_pc1969 emp_rate1969
foreach v of local vars {
    gen `v'_year = `v' * year
    local controls `v'_year 
}
* Call did_multiplegt_dyn with controls
did_multiplegt_dyn log_curr_trans_ind_gov_pc county_fips year hurricane, effects(11) placebo(11) controls(`controls') cluster(county_fips)	
\end{lstlisting}
\begin{figure}[H]
    \centering
    \includegraphics[width=0.7\textwidth]{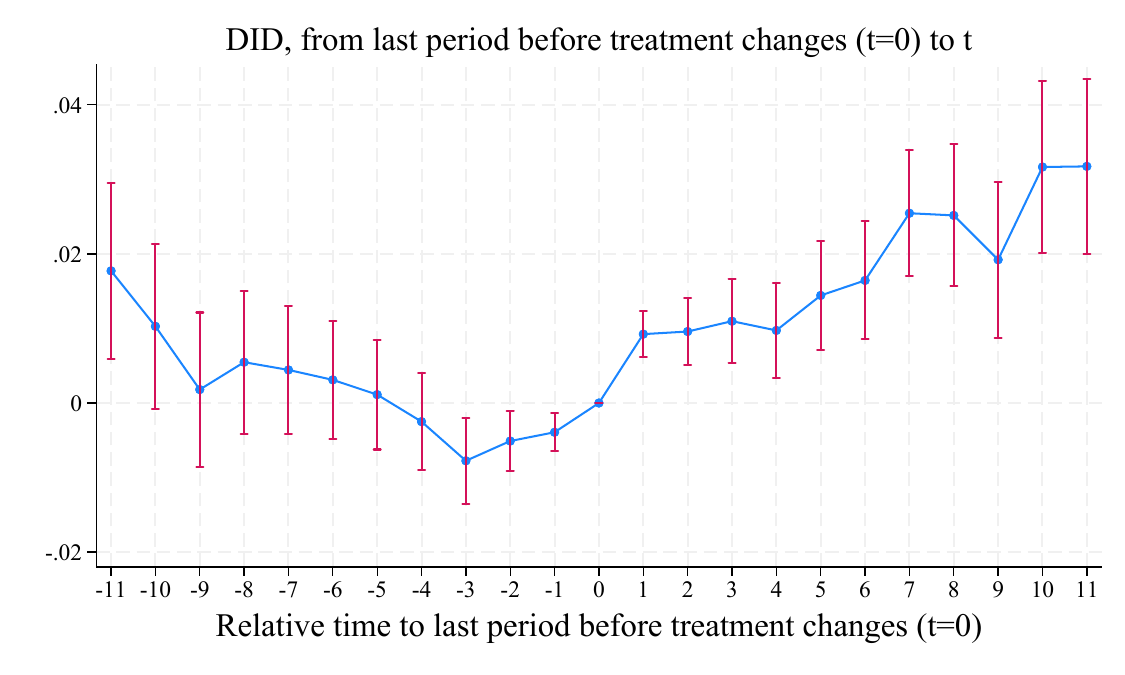}  
    \caption{\texttt{did\_multiplegt\_dyn}: Effects of Hurricane Hits on Government Transfers Per Capita, with county-level controls}
    \label{fig:yourlabel}
\end{figure}

After including those controls, the pre-trends estimators are slightly less significant, but the differential trend starting 4 periods before treatment persists. 

\subsection{\cite{east2023}: Treatment Continuously Distributed in Period 1}

\subsubsection{Research Question}

How does prenatal Medicaid expansion affect birth weight for the directly treated generation?

\subsubsection{Data}

\begin{lstlisting}
use east_et_al_2023.dta, replace
\end{lstlisting}

The data come from the publicly available dataset in the replication package of \cite{east_dataset2022}. The data is a panel of US states at the yearly level. The variable \texttt{newsimeli} is the treatment. To construct it, we start from the simulated prenatal Medicaid eligibility rate in each state and year, constructed by \cite{east2023}. This simulated eligibility rate is designed to isolate changes driven solely by Medicaid policy rules, independent of changes in states’ demographic and economic characteristics. The authors classify 28 states, in which eligibility was stagnant between 1975 and 1979, and that later experienced a large positive
shock, as their treatment group. The remaining 22 states are the control group. We construct a continuous treatment variable in the spirit of their classification. For the 22 control states, \texttt{newsimeli} is equal, throughout the study period, to the average simulated eligibility rate from 1975 to 1979. For the 28 treated states, \texttt{newsimeli}'s definition is the same before the year of the large positive shock identified by the authors. After that year, \texttt{newsimeli} is equal to the simulated eligibility rate in the year of the shock. Thus, we construct a treatment which varies across states at every time period, to reflect variation in Medicaid coverage across states. At the same time, for each state the treatment changes at most once over time: identification will only leverage the large shocks in eligibility rules identified by the authors. The outcome is \texttt{lbw\_detrend81}, the percentage of infants born in state $g$ and year $t$ who were below the low birthweight threshold, netted from a state-specific pre-treatment linear trend. 

\subsubsection{Design: the treatment takes several different values in period 1}

\begin{lstlisting}
tab newsimeli if dob_y_p==1975
\end{lstlisting}

\subsubsection{Estimate Treatement Effects with \texttt{did\_multiplegt\_dyn}}

We first use \texttt{did\_multiplegt\_dyn} to estimate non-normalized event-study effects of experiencing a large increase of prenatal Medicaid eligibility rate on states' low-birth-weight (LBW) rates. We compute five event-study estimators and five pre-trends estimators. Because the period-one treatment is continuous, we specify the \texttt{continuous(1)} option, thus assuming that states's counterfactual trends without a treatment change is linear in their baseline treatment. See Section 1.10 of the Web Appendix of \cite{de2022difference}  for further details. With the \texttt{continuous} option, the analytic ses computed by the command may not always be reliable, so we use the boostrap instead. Finally, we control for the same time-varying covariates as the authors, and we weight the estimation by the number of births in state $g$ and year $t$. We also use the \texttt{effects\_equal("all")} option to perform an F-test of the null hypothesis that all five estimated dynamic effects are equal. 

\begin{lstlisting}
global stcontrols stmarried stblack stother sthsdrop ///
sthsgrad stsomecoll pop0_4 pop5_17 pop18_24 pop25_44 ///
pop45_64 urate incpc maxafdc abortconsent abortmedr
did_multiplegt_dyn lbw_detrend81 plborn dob_y_p newsimeli, effects(5) placebo(5) continuous(1) bootstrap(50,1) controls($stcontrols) weight(births) effects_equal("all")
\end{lstlisting}

\begin{figure}[H]
    \centering
    \includegraphics[width=0.7\textwidth]{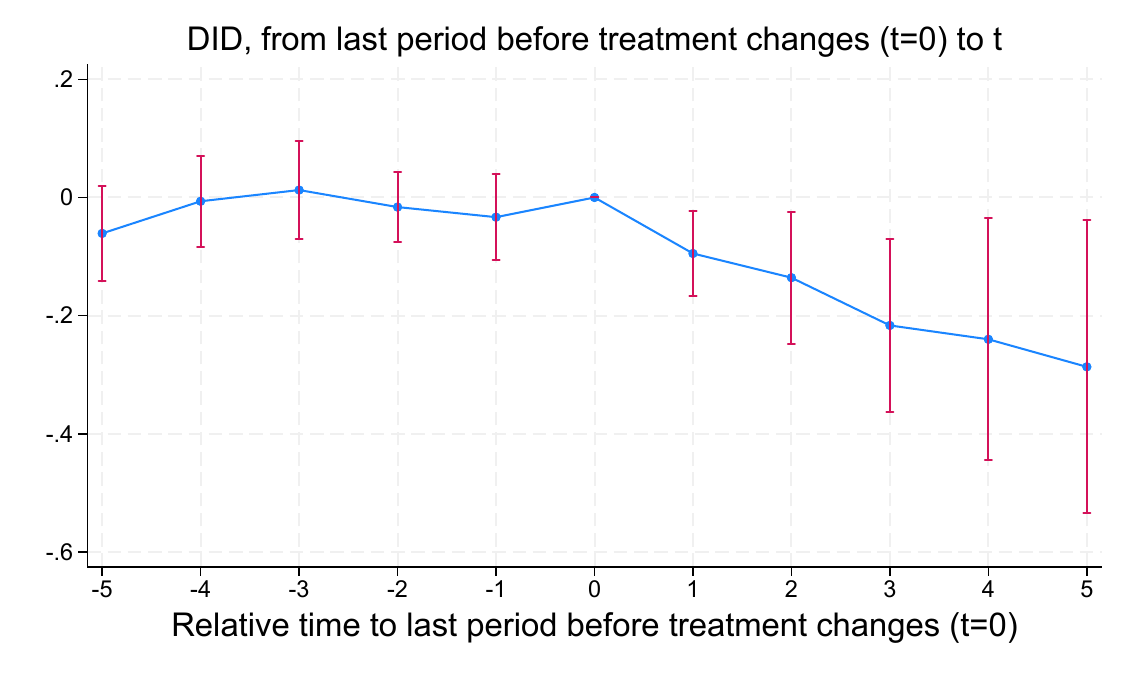}  
    \caption{Effects of a Large Increase in Prenatal Medicaid Eligibility on Low-Birth-Weight Rate}
    \label{fig:yourlabel}
\end{figure}

Results suggest that experiencing a large increase of prenatal Medicaid eligibility rate significantly decreases states' LBW rates. Pre-trend estimators are small and insignificant, thus suggesting that the no-anticipation and parallel-trends assumptions might be plausible in this application.   

\medskip
Instead of estimating a reduced-form effect of ``experiencing a large increase of prenatal Medicaid eligibility rate'', one might be interested in estimating how the LBW rate responds to a 1 percentage point increase in the prenatal Medicaid eligibility rate. For that purpose, we now use \texttt{did\_multiplegt\_dyn} to estimate normalized event-study effects.

\begin{lstlisting}
did_multiplegt_dyn lbw_detrend81 plborn dob_y_p newsimeli, effects(5) placebo(5) continuous(1) bootstrap(50,1) controls($stcontrols) weight(births) normalized normalized_weights effects_equal("all")
\end{lstlisting}

\begin{figure}[H]
    \centering
    \includegraphics[width=0.7\textwidth]{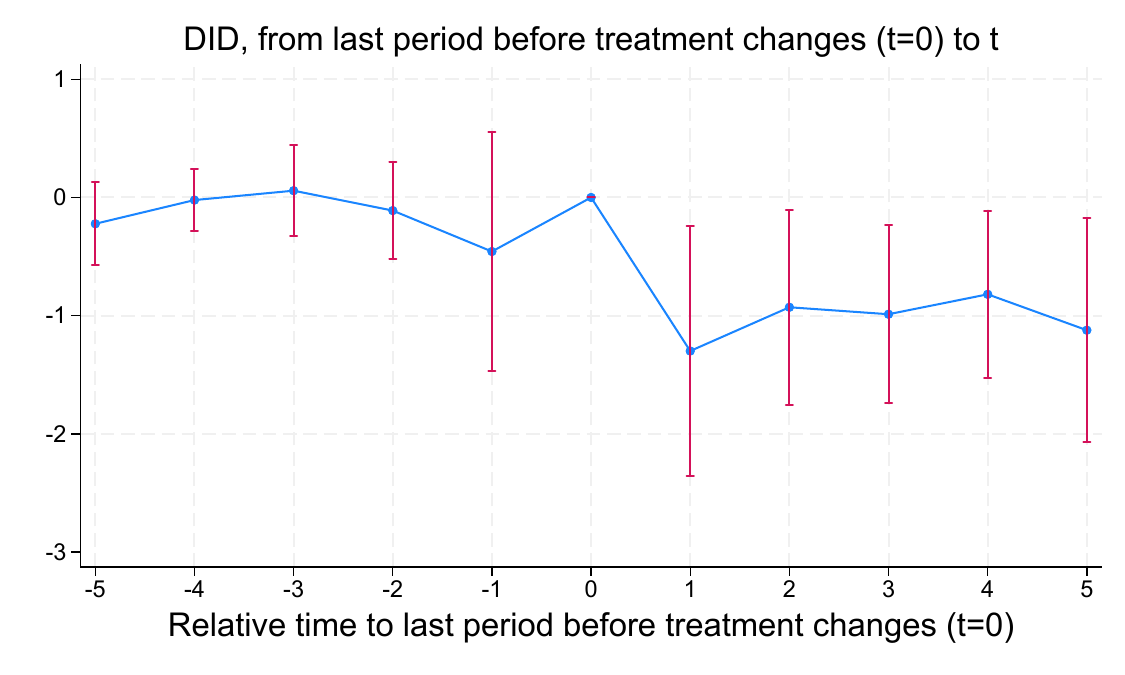}  
    \caption{Effects of a One Percentage Point Increase in Prenatal Medicaid Eligibility on Low-Birth-Weight Rate}
    \label{fig:yourlabel}
\end{figure}

The first normalized event-study estimator indicates that increasing the eligibility rate by one percentage point reduces the LBW rate by one percentage point. The second normalized event-study estimator is a weighted average, with weights 1/2 and 1/2, of the effects of the current and lagged eligibility rates on the LBW rate of state $g$ and year $t$. Again, a weighted average of the effects of increasing the current and lagged eligibility rates by one percentage point leads to a reduction of the LBW rate by one percentage point. 
The third normalized event-study estimator is a weighted average, with weights 1/3, 1/3, and 1/3, of the effects of the current and first and second lags of the eligibility rate, etc. Pre-trend estimators are normalized by the same quantity as the event-study estimators, to ensure that they can give researchers a sense of the magnitude of the potential bias in the normalized event-study estimators coming from differential trends. 

\subsection{\cite{gentzkow2011effect}: On-and-off Discrete Treatment}

\subsubsection{Research Question}

What is the effect of daily newspapers on voter turnout?

\subsubsection{Data}

\begin{lstlisting}
use gentzkowetal_didtextbook.dta, clear
\end{lstlisting}
The data come from the publicly available dataset in the replication package of \cite{gentzkow_dataset2011}. 
The data is an imbalanced panel of 1,195 US counties 
at the presidential-election-year level, from the 1872 to the 1928 election. The variable \texttt{numdailies}, the number of English-language daily newspapers in circulation in county $g$ and election-year $t$. The outcome is \texttt{prestout}, the turnout rate in county $g$ and election-year $t$.

\subsubsection{Design: On-and-off Discrete Treatment}

We first examine the values that the treatment variable \texttt{numdailies} takes in the first period of observation. 

\begin{lstlisting}
xtset cnty90 year
sort cnty90 year
bysort cnty90 (year): gen first_numdailies = numdailies[1]
tab first_numdailies
\end{lstlisting}
It takes the following values: $\{0,1,2,3,4,5,6,7,9,11,12,16,33\}$. The treatment is a multivalued discrete variable, even in the first election-year in the data.

\medskip
We then compute the first-difference of \texttt{numdailies}.

\begin{lstlisting}
gen d_numdailies = .
bysort cnty90 (year): replace d_numdailies = numdailies - numdailies[_n-1]
tab d_numdailies
\end{lstlisting}
\texttt{d\_numdailies} takes positive and negative values: counties can experience both increases and decreases in their number of newspapers.

\medskip
Finally, we compute the number of times a county experiences a change in \texttt{numdailies} over the duration of the panel.

\begin{lstlisting}
gen switch = (d_numdailies!=0&d_numdailies!=.)
bys cnty90: egen switches=total(switch)
tab switches
\end{lstlisting}
The vast majority of counties experience more than one change in \texttt{numdailies}. 19\% of counties experience 7 changes or more.

\subsubsection{Estimate Treatement Effects with \texttt{did\_multiplegt\_dyn}}

We start by computing four non-normalized event-study estimators and four pre-trends estimators.

\begin{lstlisting}
did_multiplegt_dyn prestout cnty90 year numdailies, effects(4) placebo(4) effects_equal("all")
\end{lstlisting}

\begin{figure}[H]
    \centering
    \includegraphics[width=0.7\textwidth]{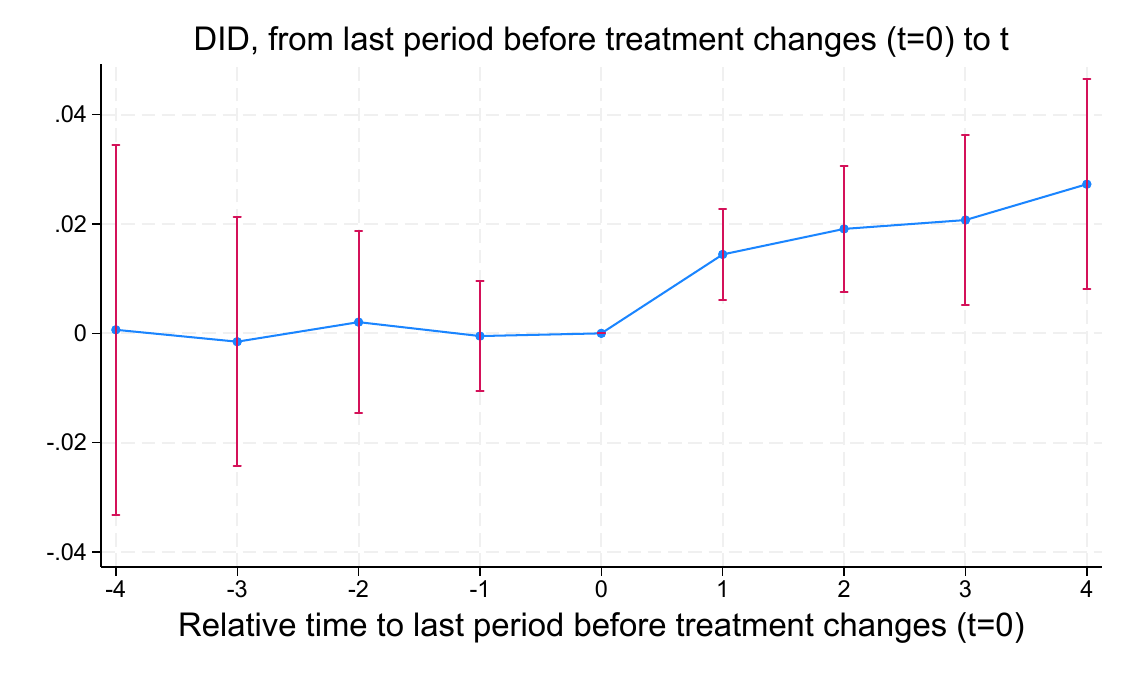}  
    \caption{\texttt{did\_multiplegt\_dyn}: Non-normalized Effects of Newspaper Entries/Exits on Voters Turnout}
    \label{fig:yourlabel}
\end{figure}

Being exposed to a weakly larger number of newspapers for one electoral cycle increases turnout, and the effect is statistically significant. That effect can be estimated for 1,119 out of the 1,195 counties in the data: 34 counties never experience a change in their number of newspapers, and 42 counties that do experience a change cannot be matched with a not-yet-switcher with the same number of newspapers at baseline. Being exposed to a weakly larger number of newspapers for two, three, and four electoral cycle also significantly increases turnout. Effects increase with exposure length, but one cannot reject the null that all effects are equal. As $\ell$ increases, effects mechanically apply to fewer and fewer counties, but the effect after four electoral cycles still applies to 917 counties. Pre-trend estimates are small and individually and jointly insignificant. However, their confidence intervals are quite large. The confidence interval of the fourth pre-trend estimator is already quite large, but that of the fifth one (not shown) is substantially larger, so we have very little power to detect differential trends over more than five election cycles. This is why we only report four placebo and four event-study estimators.

\medskip
Next, we call the \texttt{ design([float], string)} option.  This option reports switchers' period-one and subsequent treatments, thus helping the analyst understand the treatment paths whose effect is aggregated in the non-normalized event-study effects. When the number of treatment paths is large, one may specify a number $x$ included between 0 and 1 in the float argument, so that the command only reports the most common paths accounting for $x$\% of the treatment effects aggregated in the non-normalized event-study effects.

\begin{lstlisting}
did_multiplegt_dyn prestout cnty90 year numdailies, effects(2) design(0.8,console)
\end{lstlisting}

The interpretation goes as follows: 32\% of the effects aggregated by the second event-study estimator are effects of having  1 newspaper and 1 lagged newspaper instead of having 0 newspaper and 0 lagged newspaper, 18\% are effects of having 0 newspaper and 1 lagged newspaper instead of having 0 newspaper and 0 lagged newspaper, and 12\% are effects of having 2 newspapers and 1 lagged newspaper instead of having 0 newspaper and 0 lagged newspaper. 

\medskip
Next, we compute normalized event-study effects.
\begin{lstlisting}
did_multiplegt_dyn prestout cnty90 year numdailies, effects(4) placebo(4) normalized normalized_weights effects_equal("all")
\end{lstlisting}

\begin{figure}[H]
    \centering
    \includegraphics[width=0.7\textwidth]{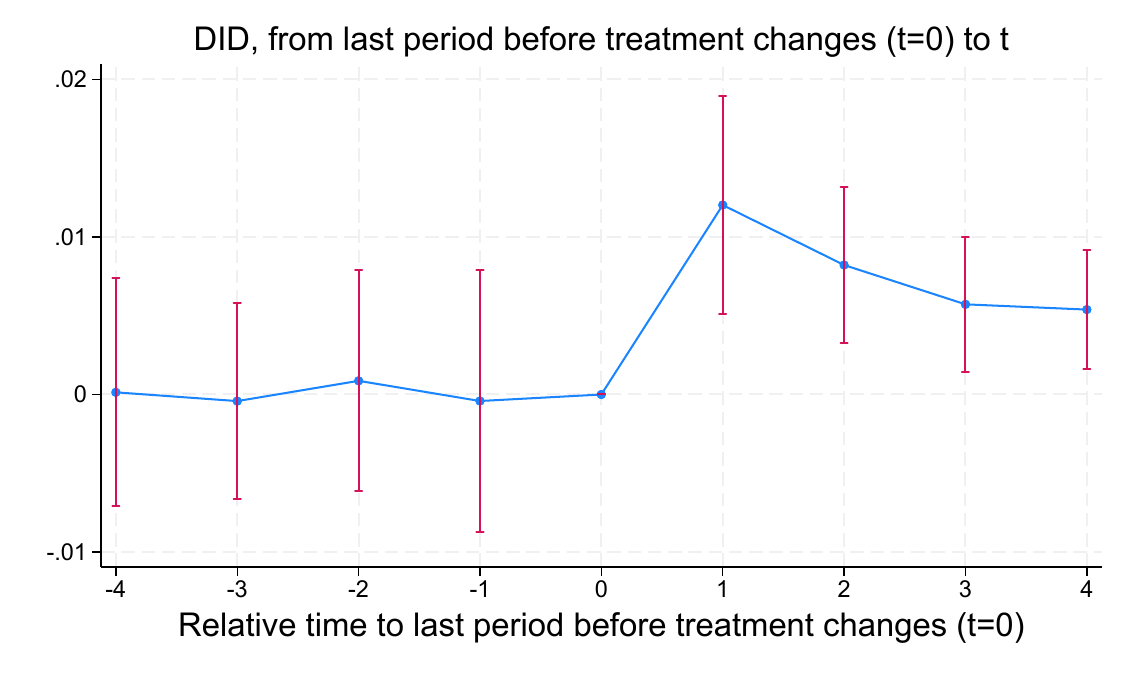}  
    \caption{\texttt{did\_multiplegt\_dyn}: Normalized Effects of Newspaper Entries/Exits on Voters Turnout}
    \label{fig:yourlabel}
\end{figure}

Normalized event-study estimates are decreasing with $\ell$, but one cannot reject the null that all effects are equal (p-value=0.17). The table ``Weights on treatment lags'', produced when the \texttt{normalized\_weights} option is specified, shows that the first event-study estimate is an effect of contemporaneous newspapers on turnout. The second normalized effect is a weighted average of the effects of contemporaneous newspapers and of the first lag of newspapers on turnout, with approximately equal weights. The third normalized effect is a weighted average of the effects of contemporaneous newspapers, the first lag of newspapers, and the second lag of newspapers, with approximately equal weights, etc.  Then, the fact that normalized event-study estimates are decreasing with $\ell$ may suggest that lagged newspapers have a smaller effect on turnout than contemporaneous newspapers.

\medskip
Finally, we run a joint test that the first lagged treatment has no effect on the outcome, and that treatment effects are constant over time. For that purpose, we estimate the first two non-normalized event-study effects, with the same \texttt{same\_switchers} option to restrict the sample to counties for which both effects can be estimated, and further restricting the sample to counties whose number of newspapers does not change in the election cycle just after their newspapers changed for the first time or whose number of newspapers has not changed yet (\texttt{year<=first\_change| same\_treat\_after\_first\_change==1}). Then, the test evaluates whether the two non-normalized effects are equal within the selected subsample.

\begin{lstlisting}
did_multiplegt_dyn prestout cnty90 year numdailies
if year<=first_change|same_treat_after_first_change==1,
effects(2) effects_equal("all") same_switchers graph_off
\end{lstlisting}

With a p-value of 0.83, the test is not rejected, thus suggesting that the first lag of newspapers does not affect current turnout.

\subsection{\cite{Hollingsworth2022}: Two Consecutive Binary and Absorbing  Treatments} 

\subsubsection{Research Question}

How do medical and recreational marijuana laws differentially affect marijuana use among adults and adolescents?

\subsubsection{Data}

\begin{lstlisting}
use hollingsworth_et_al_2022.dta, clear
\end{lstlisting}

The data come from the publicly available dataset in the replication package of \cite{hollingsworth_dataset2022}. The data is a panel of US states at the yearly level.  The treatment variables \texttt{mm} and \texttt{rm}, are respectively equal to 1 if state $g$ at year $t$ has legalized Marijuana consumption for medical and recreational purposes. The outcome variable is \texttt{ln\_mj\_use\_365}, the log of marijuana consumption in state $g$ and year $t$.

\subsubsection{Design: Two Consecutive Binary and Absorbing Treatments}

First, we verify that the two treatments are binary and absorbing.
\begin{lstlisting}
xtset state year
gen d_mm = D.mm
gen d_rm = D.rm
tab mm
tab rm
tab d_mm
tab d_rm
\end{lstlisting}
The treatments take values 0 or 1. The first-difference of the treatments take values of 0 or 1. Therefore, the two treatments are binary and absorbing.

\medskip
Next, we verify that the medical legalization treatment always happens before the recreational legalization treatment.

\begin{lstlisting}
gen mm_rm_timing = mm-rm
tab mm_rm_timing 
\end{lstlisting}
The difference between \texttt{mm} and \texttt{rm} is always positive, thus showing that the medical legalization treatment always happens before the recreational legalization treatment. 

\medskip
As we have two consecutive binary and absorbing treatments, we follow Section 8.3.4.6 of \cite{de2023credible} to separately estimate the effect of each treatment.

\subsubsection{Estimate Effects of the First Treatment with \texttt{did\_multiplegt\_dyn}}

First, we estimate the effect of medical marijuana laws. For that purpose, we just restrict  the sample to all state$\times$year $(g,t)$ such that state $g$ has not passed a recreational law yet in year $t$.
\begin{lstlisting}
did_multiplegt_dyn ln_mj_use_365 state year mm if rm==0, placebo(3) effects(3) 
\end{lstlisting}
 \begin{figure}[H]
    \centering
    \includegraphics[width=0.7\textwidth]{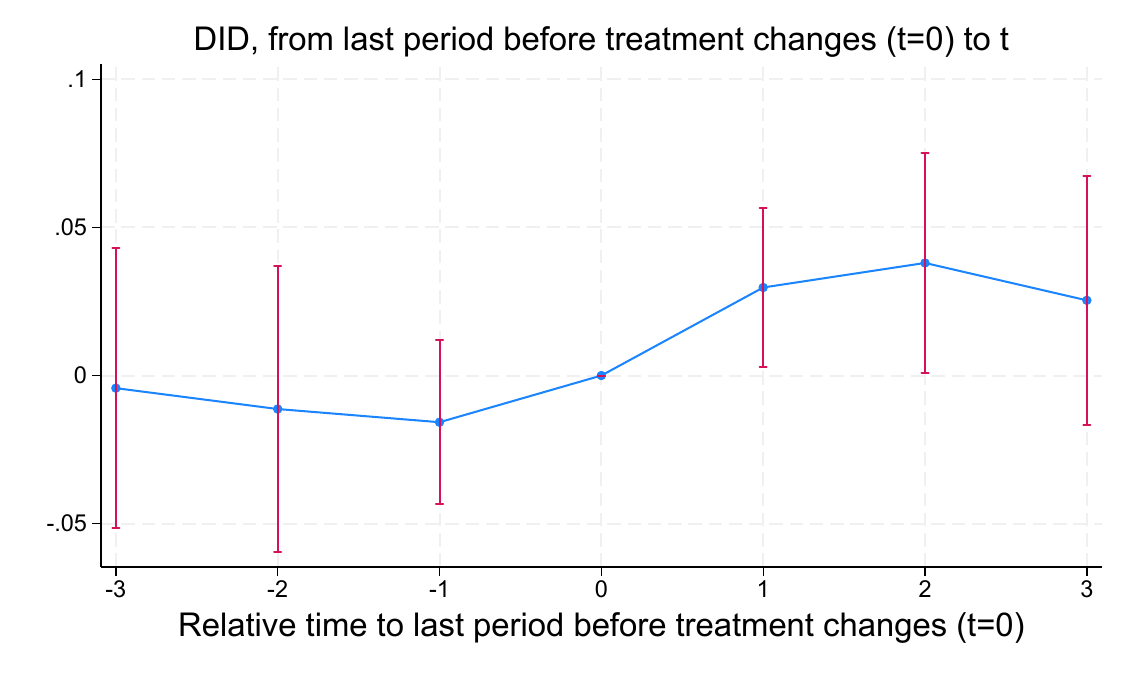}  
    \caption{\texttt{did\_multiplegt\_dyn}: Effects of Medical Marijuana Laws on Marijuana Use}
    \label{fig:yourlabel}
\end{figure}
Results suggest that medical marijuana laws increase marijuana consumption by 3-4\%, though effects are marginally significant. Pre-trends are small and insignificant, but their confidence intervals are quite large.

\subsubsection{Estimate Effects of the Second Treatment with \texttt{did\_multiplegt\_dyn}}

Next, we estimate the additional effect of a recreational marijuana law. We can estimate that effect under the assumption
that the effect of one additional period of exposure to a medical law is the same in every
group. This assumption enables us to disentangle the effect of the second treatment from the state-specific incremental impact of  the medical law treatment for an additional period. Under that assumption, estimation goes as follows. First, we restrict the sample to $(g,t)$ cells such that state $g$ has implemented a medical law at $t$. Then, we record the date when that medical law was passed, which will be inputted in the \texttt{trends\_nonparam()} option when estimating the effects of recreational laws. This ensures that the estimators will compare outcome evolutions between states that adopt the recreational law and those that do not, conditional on having adopted the medical law in the same year. 
\begin{lstlisting}
egen fg1 = min(cond(mm == 1, year, .)), by(state)
did_multiplegt_dyn ln_mj_use_365 state year rm if mm ==1, placebo(3) effects(3) trends_nonparam(fg1)
\end{lstlisting}

\begin{figure}[H]
    \centering
    \includegraphics[width=0.7\textwidth]{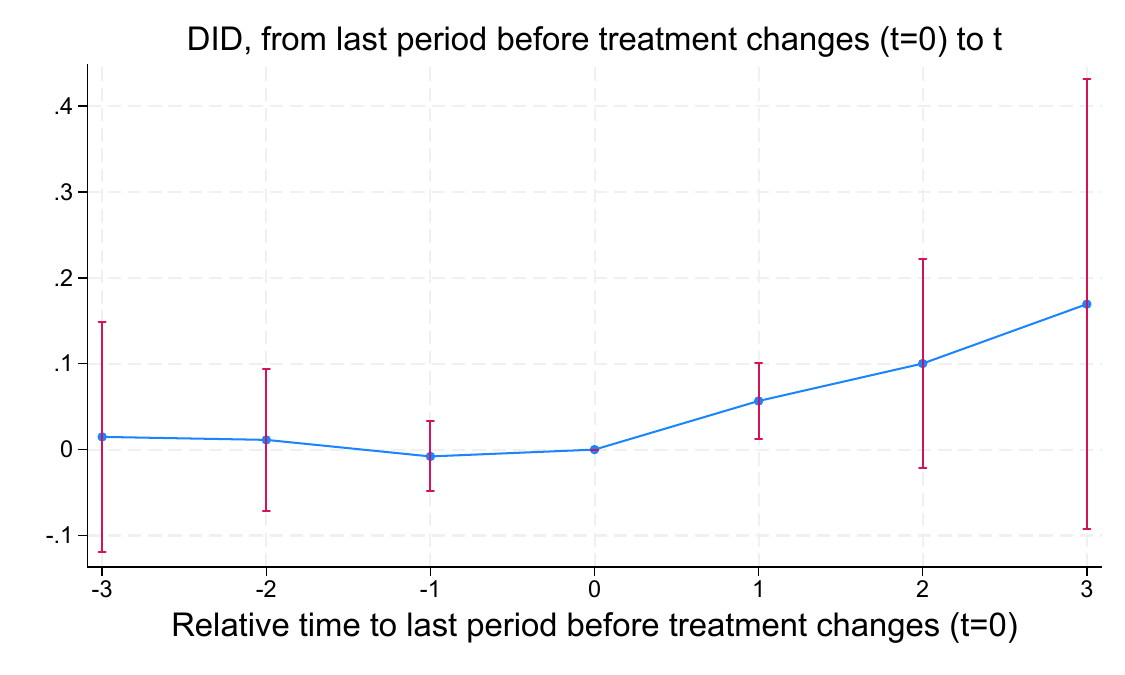}  
    \caption{\texttt{did\_multiplegt\_dyn}: Effects of Recreational Marijuana Laws on Marijuana Use.}
    \label{fig:yourlabel}
\end{figure}

Results suggest that recreational laws increase consumption by 6\% in the short run, and by 16\% in the long run, though effects are marginally significant. Effects apply to a small number of states (the effect after one year is estimated for seven states, that after two years is estimated for five states, that after three years is estimated for two states).\footnote{This is not due to controlling for the adoption date of medical marijuana laws, effects are estimated for the same number of states without that control.} Then, the asymptotic approximations underlying the command's analytic standard errors may not be reliable. Somewhat reassuringly, using the bootstrap yields similar standard errors. Pre-trends are small and insignificant, but their confidence intervals are quite large.

\bibliography{biblio}

\begin{thebibliography}{}

\bibitem[\protect\citeauthoryear{Bertrand, Duflo, and Mullainathan}{Bertrand
  et~al.}{2004}]{bertrand2004}
Bertrand, M., E.~Duflo, and S.~Mullainathan (2004).
\newblock How much should we trust differences-in-differences estimates?
\newblock {\em The Quarterly Journal of Economics\/}~{\em 119\/}(1), 249--275.

\bibitem[\protect\citeauthoryear{de~Chaisemartin and
  D'Haultf{\oe}uille}{de~Chaisemartin and D'Haultf{\oe}uille}{2020}]{dcDH2020}
de~Chaisemartin, C. and X.~D'Haultf{\oe}uille (2020).
\newblock Two-way fixed effects estimators with heterogeneous treatment
  effects.

\bibitem[\protect\citeauthoryear{de~Chaisemartin and
  D'Haultf{\oe}uille}{de~Chaisemartin and
  D'Haultf{\oe}uille}{2023a}]{dCDHtextbook}
de~Chaisemartin, C. and X.~D'Haultf{\oe}uille (2023a).
\newblock "credible answers to hard questions: Differences-in-differences for
  natural experiments.
\newblock {\em Available at SSRN\/}.

\bibitem[\protect\citeauthoryear{de~Chaisemartin and
  D'Haultf{\oe}uille}{de~Chaisemartin and
  D'Haultf{\oe}uille}{2023b}]{de2022survey}
de~Chaisemartin, C. and X.~D'Haultf{\oe}uille (2023b).
\newblock Two-way fixed effects and differences-in-differences with
  heterogeneous treatment effects: A survey.
\newblock {\em The Econometrics Journal\/}~{\em 26\/}(3), C1--C30.

\bibitem[\protect\citeauthoryear{De~Chaisemartin and
  D'Haultf{\oe}uille}{De~Chaisemartin and
  D'Haultf{\oe}uille}{2024}]{de2022intertemporal}
De~Chaisemartin, C. and X.~D'Haultf{\oe}uille (2024).
\newblock Difference-in-differences estimators of intertemporal treatment
  effects.
\newblock {\em The Review of Economics and Statistics\/}.

\bibitem[\protect\citeauthoryear{de~Chaisemartin, D'Haultf{\oe}uille, Pasquier,
  and Vazquez-Bare}{de~Chaisemartin et~al.}{2022}]{de2022difference}
de~Chaisemartin, C., X.~D'Haultf{\oe}uille, F.~Pasquier, and G.~Vazquez-Bare
  (2022).
\newblock Difference-in-differences for continuous treatments and instruments
  with stayers.
\newblock arXiv preprint arXiv:2201.06898.

\bibitem[\protect\citeauthoryear{de~Chaisemartin and
  D’Haultf{\oe}uille}{de~Chaisemartin and
  D’Haultf{\oe}uille}{2023}]{de2023credible}
de~Chaisemartin, C. and X.~D’Haultf{\oe}uille (2023).
\newblock Credible answers to hard questions: Differences-in-differences for
  natural experiment.
\newblock \url{https://papers.ssrn.com/sol3/papers.cfm?abstract_id=4487202}.

\bibitem[\protect\citeauthoryear{Deryugina}{Deryugina}{2017a}]{Deryugina2017}
Deryugina, T. (2017a, August).
\newblock The fiscal cost of hurricanes: Disaster aid versus social insurance.
\newblock {\em American Economic Journal: Economic Policy\/}~{\em 9\/}(3),
  168–98.

\bibitem[\protect\citeauthoryear{Deryugina}{Deryugina}{2017b}]{deryugina_dataset2017}
Deryugina, T. (2017b).
\newblock Replication data for: The fiscal cost of hurricanes: Disaster aid
  versus social insurance.
\newblock Nashville, TN: American Economic Association [publisher], 2017. Ann
  Arbor, MI: Inter-university Consortium for Political and Social Research
  [distributor], \url{https://doi.org/10.3886/E114620V}.

\bibitem[\protect\citeauthoryear{East, Miller, Page, and Wherry}{East
  et~al.}{2022}]{east_dataset2022}
East, C., S.~Miller, M.~Page, and L.~Wherry (2022).
\newblock Data and code for: Multi-generational impacts of childhood access to
  the safety net: Early life exposure to medicaid and the next generation’s
  health.
\newblock Nashville, TN: American Economic Association [publisher], 2022. Ann
  Arbor, MI: Inter-university Consortium for Political and Social Research
  [distributor], \url{https://doi.org/10.3886/E170801V1}.

\bibitem[\protect\citeauthoryear{East, Miller, Page, and Wherry}{East
  et~al.}{2023}]{east2023}
East, C.~N., S.~Miller, M.~Page, and L.~R. Wherry (2023).
\newblock Multigenerational impacts of childhood access to the safety net:
  Early life exposure to medicaid and the next generation’s health.
\newblock {\em American Economic Review\/}~{\em 113\/}(1), 98--135.

\bibitem[\protect\citeauthoryear{Gentzkow, Shapiro, and Sinkinson}{Gentzkow
  et~al.}{2011a}]{gentzkow2011}
Gentzkow, M., J.~M. Shapiro, and M.~Sinkinson (2011a).
\newblock The effect of newspaper entry and exit on electoral politics.
\newblock {\em American Economic Review\/}~{\em 101\/}(7), 2980--3018.

\bibitem[\protect\citeauthoryear{Gentzkow, Shapiro, and Sinkinson}{Gentzkow
  et~al.}{2011b}]{gentzkow2011effect}
Gentzkow, M., J.~M. Shapiro, and M.~Sinkinson (2011b).
\newblock The effect of newspaper entry and exit on electoral politics.
\newblock {\em American Economic Review\/}~{\em 101\/}(7), 2980--3018.

\bibitem[\protect\citeauthoryear{Gentzkow, Shapiro, and Sinkinson}{Gentzkow
  et~al.}{2011c}]{gentzkow_dataset2011}
Gentzkow, M., J.~M. Shapiro, and M.~Sinkinson (2011c).
\newblock Replication data for: The effect of newspaper entry and exit on
  electoral politics.
\newblock Nashville, TN: American Economic Association [publisher], 2011. Ann
  Arbor, MI: Inter-university Consortium for Political and Social Research
  [distributor], \url{https://doi.org/10.3886/E112472V1}.

\bibitem[\protect\citeauthoryear{Hollingsworth, Wing, and
  Bradford}{Hollingsworth et~al.}{2022a}]{hollingsworth_dataset2022}
Hollingsworth, A., C.~Wing, and A.~Bradford (2022a).
\newblock Replication package for: Comparative effects of recreational and
  medical marijuana laws on drug use among adults and adolescents.
\newblock Journal of Law and Economics. Archived at Zenodo,
  \url{https://doi.org/10.5281/zenodo.6588013}.

\bibitem[\protect\citeauthoryear{Hollingsworth, Wing, and
  Bradford}{Hollingsworth et~al.}{2022b}]{Hollingsworth2022}
Hollingsworth, A., C.~Wing, and A.~C. Bradford (2022b).
\newblock Comparative effects of recreational and medical marijuana laws on
  drug use among adults and adolescents.
\newblock {\em Journal of Law and Economics\/}~{\em 65\/}(3), 515--554.

\bibitem[\protect\citeauthoryear{Liu, Wang, and Xu}{Liu
  et~al.}{2024}]{liu2024practical}
Liu, L., Y.~Wang, and Y.~Xu (2024).
\newblock A practical guide to counterfactual estimators for causal inference
  with time-series cross-sectional data.
\newblock {\em American Journal of Political Science\/}~{\em 68\/}(1),
  160--176.

\bibitem[\protect\citeauthoryear{Robins}{Robins}{1986}]{robins1986new}
Robins, J. (1986).
\newblock A new approach to causal inference in mortality studies with a
  sustained exposure period-application to control of the healthy worker
  survivor effect.
\newblock {\em Mathematical modelling\/}~{\em 7\/}(9-12), 1393--1512.

\bibitem[\protect\citeauthoryear{Shahn}{Shahn}{2023}]{shahn2023subgroup}
Shahn, Z. (2023).
\newblock Subgroup difference in differences to identify effect modification
  without a control group.
\newblock {\em arXiv preprint arXiv:2306.11030\/}.

\bibitem[\protect\citeauthoryear{Vella and Verbeek}{Vella and
  Verbeek}{1998}]{vella1998whose}
Vella, F. and M.~Verbeek (1998).
\newblock Whose wages do unions raise? a dynamic model of unionism and wage
  rate determination for young men.
\newblock {\em Journal of Applied Econometrics\/}~{\em 13\/}(2), 163--183.

\bibitem[\protect\citeauthoryear{Wolfers}{Wolfers}{2006}]{wolfers2006did}
Wolfers, J. (2006).
\newblock Did unilateral divorce laws raise divorce rates? a reconciliation and
  new results.
\newblock {\em American Economic Review\/}~{\em 96\/}(5), 1802--1820.

\end{thebibliography}

\appendix

\section{Testing that all event-study effects are equal.}

Let $\hbeta \in \R^{L}$. Define the $(L - 1) \times 1$ vector
$$
\tbeta = \left(\hbeta_k - \frac{1}{L} \sum_{j=1}^L \hbeta_j \right)_{k = 1, ..., L - 1} 
$$
Omitting the last coefficient ensures that the components of $\tbeta$ are linearly independent. Then,
\begin{align*}
\tbeta &=
\begin{pmatrix}
1 - 1/L & -1/L & ... &  -1/L &  -1/L \\
-1/L & 1 - 1/L & ... &  -1/L &  -1/L \\
\rotatebox{90}{...} & \rotatebox{90}{...} & \rotatebox{135}{...} & \rotatebox{90}{...} & \rotatebox{90}{...} \\
-1/L &  -1/L & ... &  1 - 1/L &  -1/L \\
\end{pmatrix}
\hbeta \\
&=\underbrace{\left(\left[\boldsymbol{I}_{L-1}, \boldsymbol{0}_{L-1}\right] - \left(\frac{1}{L}\right)\boldsymbol{1}_{L-1}\boldsymbol{1}_{L}'\right)}_{D} \hbeta
\end{align*}
where $[.,.]$ is the column concatenation operator, $\boldsymbol{I}_d$ is the identity matrix of order $d$, $\boldsymbol{0}_{d}$ and $\boldsymbol{1}_{d}$ are $d \times 1$ vectors of 0 and 1.

Let $U^G_{g, \ell}$ be defined as in \cite{de2022intertemporal}.
Since $U^G_{g,\ell} = (G/N_\ell) U_{g,\ell}$, it follows from Theorem 1 therein that
\begin{align*}
\sqrt{N_\ell} \frac{DID_{\ell} - \delta_\ell}{\left(\frac{1}{N_\ell} \sum_{g = 1}^G V(U^G_{g,\ell}|D)\right)^{1/2}} 
&= \frac{DID_{\ell} - \delta_\ell}{\left(\frac{1}{N_\ell^2} \sum_{g = 1}^G V(U^G_{g,\ell}|D)\right)^{1/2}} \\
&= \frac{DID_{\ell} - \delta_\ell}{\left(\frac{1}{G^2} \sum_{g = 1}^G \frac{G^2}{N_\ell^2}V(U^G_{g,\ell}|D)\right)^{1/2}} \\
&= \sqrt{G} \frac{DID_{\ell} - \delta_\ell}{\left(\frac{1}{G} \sum_{g = 1}^G V(U_{g,\ell}|D)\right)^{1/2}} \convd N(0,1)\\
\end{align*}

Let $U_{g} = (U_{g,1}, ..., U_{g, L})$, $\hat{\delta} = (DID_1, ..., DID_L)$, $\delta = (\delta_1, ..., \delta_L)$. Using similar arguments as in the proof of Theorem 1 in \cite{de2022intertemporal}, but invoking a multivariate rather than univariate CLT, one can show that
$$
\sqrt{G} \left(\underbrace{\frac{1}{G} \sum_g V(U_g|D)}_{V} \right)^{-1/2}\left(\hat{\delta} - \delta\right) \convd N(\boldsymbol{0}_L, \boldsymbol{I}_{L})
$$
Under the null of equality of dynamic effects
$$
\sqrt{G} \left(DVD'\right)^{-1/2}D\hat{\delta} \convd N(\boldsymbol{0}_{L-1}, \boldsymbol{I}_{L-1})
$$
$$
G \hat{\delta}'D'\left(DVD'\right)^{-1}D\hat{\delta} \convd \chi^2_{L-1}
$$

\section{Conventions used by the \st{did\_multiplegt\_dyn} Stata command in the presence of missing treatments}

The estimators computed by the command crucially rely on the full knowledge of a group's sequence of treatments. Of particular importance are a group's baseline treatment $D_{g,1}$, and the first period at which a group's treatment changes $F_g$. While the definitions of  $D_{g,1}$ and $F_g$ are uncontroversial when groups' treatments are observed at every time period, those definitions can become more controversial when the treatments of some groups are unobserved at some time periods. Then, \st{did\_multiplegt\_dyn} redefines $D_{g,1}$ as a group's treatment at the first time period where that treatment is observed, and $F_g$ as the first period where the group is observed with a different treatment value.   \st{did\_multiplegt\_dyn} uses the following rules to deal with missing treatments. 

\paragraph{Conservative option to deal with missing treatments.}
If the \\
\st{drop\_if\_d\_miss\_before\_first\_switch} option is specified, \st{did\_multiplegt\_dyn} adopts a conservative approach, and drops all $(g,t)$ cells such that there is a $t_0\leq t$ such that $g$'s treatment is missing at $t_0$, $g$'s treatment has not changed yet at $t_0$, and $g$'s outcome has been observed at least once at or before $t_0$. In such situations, we cannot know for sure when $g$'s first treatment change took place. 

\paragraph{Liberal option to deal with missing treatments: groups that experience at least one treatment change.}
By default, the command adopts a more liberal approach to deal with missing treatments. For groups that experience at least one treatment change, let $FD_g$ be the first date at which their treatment is observed, let $LDBF_g$ be the last date before their first treatment change when their treatment is observed, let $F_g$ denote the first date when they are observed with a different treatment than that they had at $FD_g$, and let $LD_g$ be the last period when their treatment is observed.  We have $FD_g\leq LDBF_g\leq F_g\leq LD_g$, and the command deals with missing treatments depending on when they occur with respect to those four dates:
\begin{enumerate}
\item Before $FD_g$, the command considers that $g$ joins the panel at $FD_g$: any non-missing outcome before $FD_g$ is replaced as missing, but all non-missing outcomes after $FD_g$ are kept. For groups such that their outcome is observed at least one date before $FD_g$, this is a liberal convention: those groups exist before $FD_g$, so one could argue that their baseline treatment is missing. Users may drop those groups from the analysis, using the \st{drop\_if\_d\_miss\_before\_first\_switch} option.
\item If a treatment is missing at $FD_g<t<LDBF_g$, it is replaced by $g$'s baseline treatment. In a binary and staggered design, this amounts to imputing the missing treatment of a group later observed with a treatment of zero by a zero, an imputation justified by the design. Outside of binary and staggered designs, this may be a liberal imputation, which can be overruled by using the \st{drop\_if\_d\_miss\_before\_first\_switch} option.
\item If $LDBF_g<F_g-1$, $g$'s treatment is missing at $F_g-1$, the period just before it is observed with a treatment different from its baseline treatment. Then, we cannot know the exact date when $g$'s treatment has changed for the first time, even in a binary and staggered design. Therefore, we set $g$'s outcome at missing starting at $LDBF_g+1$.
\item If the treatment is missing at $t>F_g$, we replace it by $D_{g,F_g}$, $g$'s treatment at the date of its  first treatment change. In designs where groups' treatment can change at most once, this imputation is justified by the design. In other designs, this imputation may be liberal, but it is innocuous for the computation of the reduced-form effects $\DID_\ell$, so we do not give users the option to overrule it.
\end{enumerate}

\paragraph{Liberal option to deal with missing treatments: groups whose treatment never changes.} 
For those groups, we just have $FD_g\leq LD_g$, and we will deal with missing treatments depending on when they occur with respect to those two dates.
\begin{enumerate}
\item Before $FD_g$, the command again considers that $g$ joins the panel at $FD_g$: any non-missing outcome before $FD_g$ is replaced as missing, but all non-missing outcomes after $FD_g$ are kept. This is again a liberal convention that can be overruled using the \st{drop\_if\_d\_miss\_before\_first\_switch} option.
\item Treatments missing at $FD_g<t<LD_g$ are replaced by $D_{g,1}$. In a binary and staggered design, this imputation is justified by the design. Outside of binary and staggered designs, this may be a liberal imputation, which can be overruled by using the \st{drop\_if\_d\_miss\_before\_first\_switch} option.
\item For all $t>LD_g$, $g$'s outcomes are replaced by missing. Even in a binary and staggered design, we cannot infer $g$'s  treatment at $t>LD_g$: $g$ may have gotten treated at those dates and one may just not observe it.
\end{enumerate}

\section{Comparing the \st{did\_multiplegt} and \st{did\_multiplegt\_dyn} Stata commands}

\subsection{Main differences between the two commands}

\paragraph{Estimators computed.} Like \st{did\_multiplegt\_dyn}, the \st{did\_multiplegt} package also computes the event-study estimators proposed by \cite{de2022intertemporal}, when the \st{robust\_dynamic} option is specified. \st{did\_multiplegt\_dyn} supersedes 
\st{did\_multiplegt} to compute those estimators. When the \st{robust\_dynamic} option is not specified, \st{did\_multiplegt} computes the $\DID_M$ estimator proposed by \cite{dcDH2020}. That estimator is not computed by \st{did\_multiplegt\_dyn}:  \st{did\_multiplegt} remains the primary tool to compute it.

\paragraph{Syntax.} Like \st{did\_multiplegt}, the basic syntax of \st{did\_multiplegt\_dyn} is \\ 
\st{did\_multiplegt\_dyn Y G T D},\\ 
where \st{Y} is the outcome variable, \st{G} the outcome variable, \st{T} the time variable, and \st{D} the treatment variable.

\paragraph{Variance estimation.} The main difference between \st{did\_multiplegt} and \st{did\_multiplegt\_dyn} is that the former relies on the bootstrap for variance estimation, while the later relies on analytic variance formulas. Accordingly, \st{did\_multiplegt\_dyn} is much faster than \st{did\_multiplegt}, and it does not need the \st{breps(\#)} option.

\paragraph{Notational differences.}
The notations used by \st{did\_multiplegt} and \st{did\_multiplegt\_dyn} are slightly different: \st{effect\_0} in \st{did\_multiplegt} is \st{effect\_1} in \st{did\_multiplegt\_dyn}, \st{effect\_1} in \st{did\_multiplegt} is \st{effect\_2} in \st{did\_multiplegt\_dyn}, etc... To estimate more than one event-study effect, one has to request the \st{effects} option with \st{did\_multiplegt\_dyn}. That option was called \st{dynamic} in the \st{did\_multiplegt} command. 

\paragraph{Clustering by default.} The package \st{did\_multiplegt\_dyn} clusters by default standard errors at the level of the group variable. Standard errors can be clustered at a more aggregated level, but they cannot be clustered at a more disaggregated level. \st{did\_multiplegt} clustered by default at the group$\times$time level. 

\paragraph{Options.} Some of the options of \st{did\_multiplegt} do not exist for \st{did\_multiplegt\_dyn}. This includes the \st{recat\_treatment} and \st{threshold\_stable\_treatment} options. On the other hand, \st{did\_multiplegt\_dyn} has some new options, to compute the normalized event-study estimators proposed in the latest version of \cite{de2022intertemporal}, and to test for heterogeneous treatment effects. 

\subsection{Instances where \st{did\_multiplegt} and \st{did\_multiplegt\_dyn} may output different results.}

The two commands do not output different event-study and placebo effects when applied to a balanced panel of groups, the \st{dont\_drop\_larger\_lower} option is not specified in \st{did\_multiplegt\_dyn} while the \st{drop\_larger\_lower} option is specified in \st{did\_multiplegt}, and the \st{trends\_nonparam} option is not specified. 

\paragraph{Imbalanced panel: missing treatments.} The two packages might output different results when applied to a panel of groups that is imbalanced because some groups' treatments are missing at some dates. The two commands use different conventions to account for missing treatments. The conventions used by  \st{did\_multiplegt\_dyn} are explained in great details in Appendix A. The conventions used by  \st{did\_multiplegt} are not as explicitly thought out as those used by \st{did\_multiplegt\_dyn}: the latter command deals more transparently with missing treatments. 

\paragraph{Imbalanced panel: missing outcomes.}
When the group-level panel is imbalanced because some groups have missing outcomes at some time periods, the event-study effects outputed by \st{did\_multiplegt\_dyn} and \st{did\_multiplegt} are similar, but the placebo effects can differ. When estimating a placebo effect $\DID^{pl}_\ell$,  \st{did\_multiplegt\_dyn} takes into account only the switchers which were used in the estimation of the corresponding dynamic effect $\DID_\ell$. \st{did\_multiplegt}, on the other hand, includes some switchers for which that effect could not be computed, because their $F_g-1$ or their $F_g-1+\ell$ outcome is missing. 

\paragraph{Switchers with no first stage.}
Switchers whose average treatment after they switch is exactly equal to their baseline treatment, herafter referred to as switchers with no first-stage, are dropped by \st{did\_multiplegt\_dyn}, not by  \st{did\_multiplegt}. This can lead to differences between the event-study and placebo effects outputed by the two commands. Switchers with no first stage can only arise when the treatment is non-binary. Such switchers are dropped by both commands when the \st{dont\_drop\_larger\_lower} option is not specified in \st{did\_multiplegt\_dyn} while the \st{drop\_larger\_lower} option is specified in \st{did\_multiplegt}, so the event-study and placebo effects outputed by the two commands are the same when this option is specified, even in the presence of switchers with no first-stage.

\paragraph{Missing controls.}
\st{did\_multiplegt} and \st{did\_multiplegt\_dyn} can also output  different results when the \st{controls} option is specified, and some covariates are missing for some $(g,t)$-cells. When a switcher's covariate is missing at $t$,  \st{did\_multiplegt} drops that switcher from the estimation at all subsequent time periods. \st{did\_multiplegt\_dyn}, on the other hand, does not drop that switcher from the estimation at all subsequent time periods. 

\paragraph{\st{trends\_nonparam}.}
When the \st{trends\_nonparam(\it{varlist})} option is specified, \st{did\_multiplegt\_dyn} drops switchers with no control group with the same value of \st{\it{varlist}}. \st{did\_multiplegt} only drops those switchers if its option \st{always\_trends\_nonparam} is specified. Thus, with the \st{trends\_non\_param} option, the two commands can output different results if \st{did\_multiplegt} is ran without the \st{always\_trends\_nonparam} option. 

\paragraph{Number of observations used to compute the average effect.} The number of observations outputed by \st{did\_multiplegt} was sometimes incorrect. The number outputed by \st{did\_multiplegt\_dyn} is correct.

\end{document}